\documentclass[sigconf]{acmart}

\copyrightyear{2026}
\acmYear{2026}
\setcopyright{rightsretained}
\acmConference[ICSE '26]{2026 IEEE/ACM 48th International Conference on Software Engineering}{April 12--18, 2026}{Rio de Janeiro, Brazil}
\acmBooktitle{2026 IEEE/ACM 48th International Conference on Software Engineering (ICSE '26), April 12--18, 2026, Rio de Janeiro, Brazil}
\acmPrice{}
\acmDOI{10.1145/3744916.3773149}
\acmISBN{979-8-4007-2025-3/26/04}

\newif\ifDEBUG
\DEBUGfalse
\DEBUGtrue

\newif\ifBLINDED
\BLINDEDtrue

\newif\ifAPPENDIX
\APPENDIXfalse

\newif\ifARXIV
\ARXIVtrue
\ARXIVfalse

\newcommand{\ie}{\textit{i.e.,\ }}
\newcommand{\eg}{\textit{e.g.,\ }}
\newcommand{\etal}{\textit{et al.\ }}

\usepackage{mdframed}
\mdfsetup{skipabove=0.5\topskip,skipbelow=0.5\topskip,align=center}

\usepackage{amsthm}
\newtheorem{thm}{Theorem}

\usepackage{enumitem}
\setlist[itemize]{leftmargin=*,noitemsep,topsep=0pt}
\setlist[enumerate]{leftmargin=*}

\usepackage{todonotes}
\usepackage{soul}

\ifDEBUG
    \newcommand{\DA}[1]{\todo[color=white,inline]{DA:#1}}
    \newcommand{\DANew}[1]{\todo[color=white,inline]{DA:#1}}
    \newcommand{\JD}[1]{\todo[color=yellow,inline]{JD:#1}}
    
    \newcommand{\TS}[1]{\textcolor{purple}{#1}}

    \newcommand{\TODO}[1]{\hl{#1}}
\else
    \newcommand{\DA}[1]{}
    \newcommand{\JD}[1]{}
    \newcommand{\WJ}[1]{}
    
    \newcommand{\TODO}[1]{}
\fi

\usepackage{booktabs} 
\usepackage{colortbl}
\usepackage{amsmath}
\usepackage{amsfonts}
\usepackage{algpseudocode}
\usepackage[normalem]{ulem} 
\usepackage{xspace}
\usepackage{multirow} 
\usepackage{balance} 
\usepackage{fancyvrb} 
\usepackage[export]{adjustbox} 
\usepackage{subcaption} 
\usepackage[font=small,skip=5pt]{caption}
\usepackage{url}
\usepackage{xcolor}
\usepackage{tcolorbox}
\usepackage[commandnameprefix=ifneeded, final]{changes}

\usepackage{changes}

\usepackage{tikz} 
\usetikzlibrary{automata} 
\usetikzlibrary{positioning} 
\usetikzlibrary{arrows} 
\usetikzlibrary{calc} 
\usetikzlibrary{patterns} 
\usetikzlibrary{snakes} 

\usepackage[group-separator={,}]{siunitx}

\VerbatimFootnotes 

\usepackage{cleveref}
\crefformat{section}{\S#2#1#3}
\crefname{figure}{Figure}{Figures}
\crefname{table}{Table}{Tables}
\crefname{listing}{Listing}{Listings}
\crefname{theorem}{Theorem}{Theorems}
\crefname{thm}{Theorem}{Theorems}
\crefname{lemma}{Lemma}{Lemmata}
\crefname{equation}{Eqt.}{Eqts.}
\crefformat{Grammar}{Grammar #1}

\usepackage{listings}

\newcommand{\myparagraph}[1]{\vspace{0.08cm}\noindent\hspace{0.1cm} \underline{#1:}}

\newenvironment{RQList}{
   \setlength{\topsep}{0pt}
   \setlength{\partopsep}{0pt}
   \setlength{\parskip}{0pt}
   \begin{description}[style=unboxed]
   \setlength{\leftmargin}{0.80in}
   \setlength{\parsep}{0pt}
   \setlength{\parskip}{0pt}
   \setlength{\itemsep}{0pt}
   }
   {\end{description}}

\newcommand{\myinlinequote}[1]{\emph{``#1''}}

\newenvironment{myquote}%
  {\list{}{\leftmargin=0.12in\rightmargin=0.1in}\item[]}%
  {\endlist}
\newcommand{\blockquote}[1]{\begin{myquote}``\textit{#1}''\end{myquote}}

\usepackage{newfloat}
\usepackage{syntax}
\usepackage{framed}

\DeclareFloatingEnvironment[
  fileext   = logr,
  listname  = {List of Grammars},
  name      = Grammar,
  placement = htp
]{Grammar}

\usepackage[ruled,vlined]{algorithm2e}

\usepackage{nicefrac}
\usepackage{graphicx}
\usepackage{caption}
\usepackage{subcaption}
\usepackage{float}
\usepackage{multirow}
\usepackage{tabularx}
\usepackage{ulem}

\usepackage{xspace}

\newcommand{\SubjectOne}{\emph{S1}\xspace}
\newcommand{\SubjectTwo}{\emph{S2}\xspace}
\newcommand{\SubjectThree}{\emph{S3}\xspace}
\newcommand{\SubjectFour}{\emph{S4}\xspace}
\newcommand{\SubjectFive}{\emph{S5}\xspace}
\newcommand{\SubjectSix}{\emph{S6}\xspace}
\newcommand{\SubjectSeven}{\emph{S7}\xspace}
\newcommand{\SubjectEight}{\emph{S8}\xspace}
\newcommand{\SubjectNine}{\emph{S9}\xspace}
\newcommand{\SubjectTen}{\emph{S10}\xspace}

\newcommand{\PilotSubjectOne}{\emph{PS1}\xspace}
\newcommand{\PilotSubjectTwo}{\emph{PS2}\xspace}
\newcommand{\PilotSubjectThree}{\emph{PS3}\xspace}
\newcommand{\PilotSubjectFour}{\emph{PS4}\xspace}
\newcommand{\PilotTeam}{\emph{PTeam}\xspace}

\begin{document}

\title[Learning From Software Failures: A Case Study at a National Space Research Center]{Learning From Software Failures:\\A Case Study at a National Space Research Center}

\author{Dharun Anandayuvaraj}
\affiliation{%
  \institution{Purdue University}
  \country{USA}
}
\email{dananday@purdue.edu}

\author{Tanmay Singla}
\affiliation{%
  \institution{Purdue University}
  \country{USA}
}
\email{singlat@purdue.edu}

\author{Zain A. H. Hammadeh}
\affiliation{%
  \institution{German Aerospace Center (DLR)}
  \country{Germany}
}
\email{zain.hajhammadeh@dlr.de}

\author{Andreas Lund}
\affiliation{%
  \institution{German Aerospace Center (DLR)}
  \country{Germany}
}
\email{Andreas.Lund@dlr.de}

\author{Alexandra Holloway}
\affiliation{%
  \institution{Jet Propulsion Laboratory, California Institute of Technology}
  \country{USA}
}
\email{alexandra.holloway@jpl.nasa.gov}

\author{James C. Davis}
\affiliation{%
  \institution{Purdue University}
  \country{USA}
}
\email{davisjam@purdue.edu}


\begin{abstract}

Software failures can have significant consequences, making learning from failures a critical aspect of software engineering.
While software organizations are recommended to conduct postmortems to learn from failures, the effectiveness and adoption of these practices vary widely. 
Understanding how engineers gather, document, share, and apply lessons from failures is essential for improving software reliability and preventing recurring failures. 
High-reliability organizations (HROs) often develop software systems where failures carry catastrophic risks, requiring continuous learning practices to ensure reliability. 
These organizations provide a valuable setting to examine practices and challenges for learning from software failures. 
Such insight could help develop processes and tools to improve reliability and to prevent recurring failures. 
However, we lack in-depth industry perspectives on the practices and challenges of learning from failures.

To address this gap, we conducted a case study through 10 in-depth interviews with research software engineers at a national space research center. 
We examine how they \added{learn from failures: how they} gather, document,  share, and apply lessons learned.
To assess the transferability of our findings, we include data from 5 additional interviews at other HROs.
Our findings provide insight on how software engineers learn from failures in practice.
To summarize our findings: 
(1) failure learning is informal, ad-hoc, and inconsistently integrated into SDLC;
(2) recurring failures persist due to the absence of structured processes; and  
(3) key challenges, including time constraints, knowledge loss due to team turnover \& fragmented documentation, and weak process enforcement, undermine efforts to systematically learn from failures.  
Our findings contribute to a deeper understanding of how software engineers learn from failures and offer guidance for improving failure management practices.

\end{abstract}


\maketitle


\section{Introduction}

Engineers expect some defects \cite{kuutilaTimePressureSoftware2020, costello1984software} but strive to eliminate severe ones that cause incidents: undesired, unplanned, software-induced events causing substantial loss \cite{leveson1995safeware,avizienisBasicConceptsTaxonomy2004}. The presence of these defects, whether caught internally or as incidents, signifies a \underline{failure} in the software engineering process. Understanding these failures is vital for failure mode analysis \cite{reiferSoftwareFailureModes1979,song2012applying,ishimatsu2010modeling} to help identify and prioritize risks \cite{fairbanksJustEnoughSoftware2010}.

A core engineering principle is the systematic analysis of failures to prevent their recurrence~\cite{petroskiDesignParadigmsCase1994, dalcherFallingPartGrowing1994}. 
This approach has been effective across various engineering disciplines, contributing to the low failure rates of \eg buildings, aircraft, and nuclear reactors.
Postmortem and retrospective practices, where engineers systematically analyze failures and gather lessons learned~\cite{collier1996defined}, are among the recommended methods for institutionalizing learning from software failures.
However, despite the recognized importance of these practices, they are rarely followed consistently as part of the software engineering process within organizations~\cite{vieiraTechnicalManagerialDifficulties2019, kasiPostMortemParadox2008, dingsoyrPostmortemReviewsPurpose2005, glassProjectRetrospectivesWhy2002, keeganQuantityQualityProjectBased2001}. 
We believe that one contributing factor to the persistent high rate of software failures is the limited adoption and effectiveness of practices to learn from failures~\cite{anandayuvarajReflectingRecurringFailures2023}.
Although existing works have provided general methods for conducting postmortems~\cite{dingsoyrPostmortemReviewsPurpose2005, tiedemanPostmortemsmethodologyExperiences1990, collier1996defined, myllyaho2004review, stalhanePostMortemAssessment2003, dingsoyrAugmentingExperienceReports2001} and observations on how postmortem practices have been implemented~\cite{desouzaExperiencesConductingProject2005c, washburn2016went, birkPostmortemNeverLeave2002, kasiPostMortemParadox2008, lyytinenLearningFailureInformation1999, dingsoyrOrganizationalLearningProject2007,  vieiraTechnicalManagerialDifficulties2019}, we lack empirical insight on the practices and challenges faced by engineers to \replaced{learn from failures: how they gather, document,  share, and apply lessons learned.}{study, document, reflect on, share, and learn from failure knowledge}.

\added{Such insights are particularly valuable to study in} high-reliability organizations (HROs), such as those in aerospace, healthcare, and finance, because their software systems operate under strict safety, security, and performance constraints to maintain reliability~\cite{christiansonBecomingHighReliability2011}. 
In these environments, software failures can have severe consequences, as demonstrated by recent software incidents with catastrophic impacts~\cite{anandayuvaraj2024fail, koThirtyYearsSoftware2014, wongBeMoreFamiliar2017, wongRoleSoftwareRecent2009, zollersNoMoreSoft2005, NAP11923}, ranging from financial losses to threats to human life.
As highlighted by the National Academy of Engineers~\cite{NAP11923}, software engineering failures are widespread, with annual costs estimated in the trillions of dollars~\cite{krigsmanAnnualCostIT2009b}, emphasizing the need to reduce software failures. 
Consequently, organizations \textit{should} invest resources into software failure analysis and learning processes to reduce recurrence and improve reliability.

Our work presents a case study to examine how engineers at an HRO learn from failures. 
We conducted 10 semi-structured interviews with software engineering researchers at a national space research center. 
To assess the transferability of our findings, we include 5 additional interviews with engineers from four other HROs.
The interviews focused on participants' practices and challenges in gathering, documenting, sharing, and applying lessons learned for ongoing improvement.

Our results provide insight into how software engineers in HROs learn from failures. 
Although HROs are expected to excel at failure management, our study reveals that structured processes for learning from failures is often informal and \textit{ad hoc}.
Systemic barriers, such as time constraints, documentation gaps, and team turnover, limit effective learning. 
As a result, recurring failures persist despite general process adherence.
In summary, our contributions are:
\noindent
\begin{enumerate}
	\item We characterize the practices used by engineers in an HRO to gather, document, share, and apply lessons from failures.
    \item We identify key challenges -- technical, organizational, and cultural -- that impact learning from failures.
    \item We propose recommendations and research directions to enable learning from failures. 
\end{enumerate}

\noindent
\textbf{\ul{Significance}}:
Our study provides insights on the (\textit{ad hoc}) practices currently used in some HROs to learn from failures. 
Our study highlights organizational, technical, and cultural obstacles to more effective failure management. 
We also provide recommendations for practitioners to institutionalize structured practices to learn from failures.
We hope these insights facilitate continuous improvement and learning, with the goal of reducing recurring failures in high-reliability engineering contexts.

\section{Background and Related Work} \label{sec:Background}

We discuss
  the phenomenon of
  \textit{learning from failures} in traditional engineering disciplines (\cref{sec:Background-Engineering})
  and
  in software engineering (\cref{sec:Background-SoftwareEngineering}).

\begin{tcolorbox}[title=Definition: Learning from Failures,boxsep=1pt, left=1pt, right=1pt, top=1pt, bottom=1pt]
\textbf{Learning from failures} refers to practices by which practitioners gather, document,  share, and apply lessons from failures~\cite{dingsoyrPostmortemReviewsPurpose2005, kasiPostMortemParadox2008, christiansonBecomingHighReliability2011}.
\end{tcolorbox}


\subsection{Learning from Failures in Engineering} \label{sec:Background-Engineering}

An essential principle of engineering is to analyze failures, learn from them, and implement mitigation to reduce risk of future recurrence ~\cite{petroskiDesignParadigmsCase1994, reasonHumanError1990, dalcherFallingPartGrowing1994}.
This principle has been successful in many engineering disciplines, where lessons from failures has contributed to the low failure rates of, e.g., medical devices, aircraft, railways, and nuclear reactors. 
For example, there are databases of failures in
  medical devices~\cite{commissionerMedWatchFDASafety2022}, 
  aviation~\cite{ASRSAviationSafety},
  aerospace~\cite{NASA2023LessonsLearned},
  railways~\cite{RailAccidentInvestigation2017},
  and
  chemicals~\cite{InvestigationsCSB}, to enable practitioners in those fields to learn from past failures.
Methodologies like Six Sigma are used by manufacturing~\cite{gijoApplicationSixSigma2011}, chemical~\cite{falconImprovingEnergyEfficiency2012}, civil~\cite{moralesSixSigmaImprovement2016}, and automotive~\cite{antonyApplicationSixSigma2005} engineering fields to systematically analyze errors and improve processes~\cite{antonyDesignSixSigma2002}.

Non-software engineering disciplines emphasize learning from past failures to identify faults and mitigate them in earlier development stages (\eg FMEA~\cite{goddard2000software}), because the cost of addressing faults post-implementation can be significant~\cite{zage2003analysis,McConnell2001once,stecklein2004error}.
The physical constraints of traditional engineering fields often lead to Waterfall-type process. 
Repairing errors in such processes is costly~\cite{royce1970managing}.

\subsection{Learning from Failures in Software Eng.} \label{sec:Background-SoftwareEngineering}

\begin{figure}[!ht]

    \centering
    {\includegraphics[width=1\linewidth]{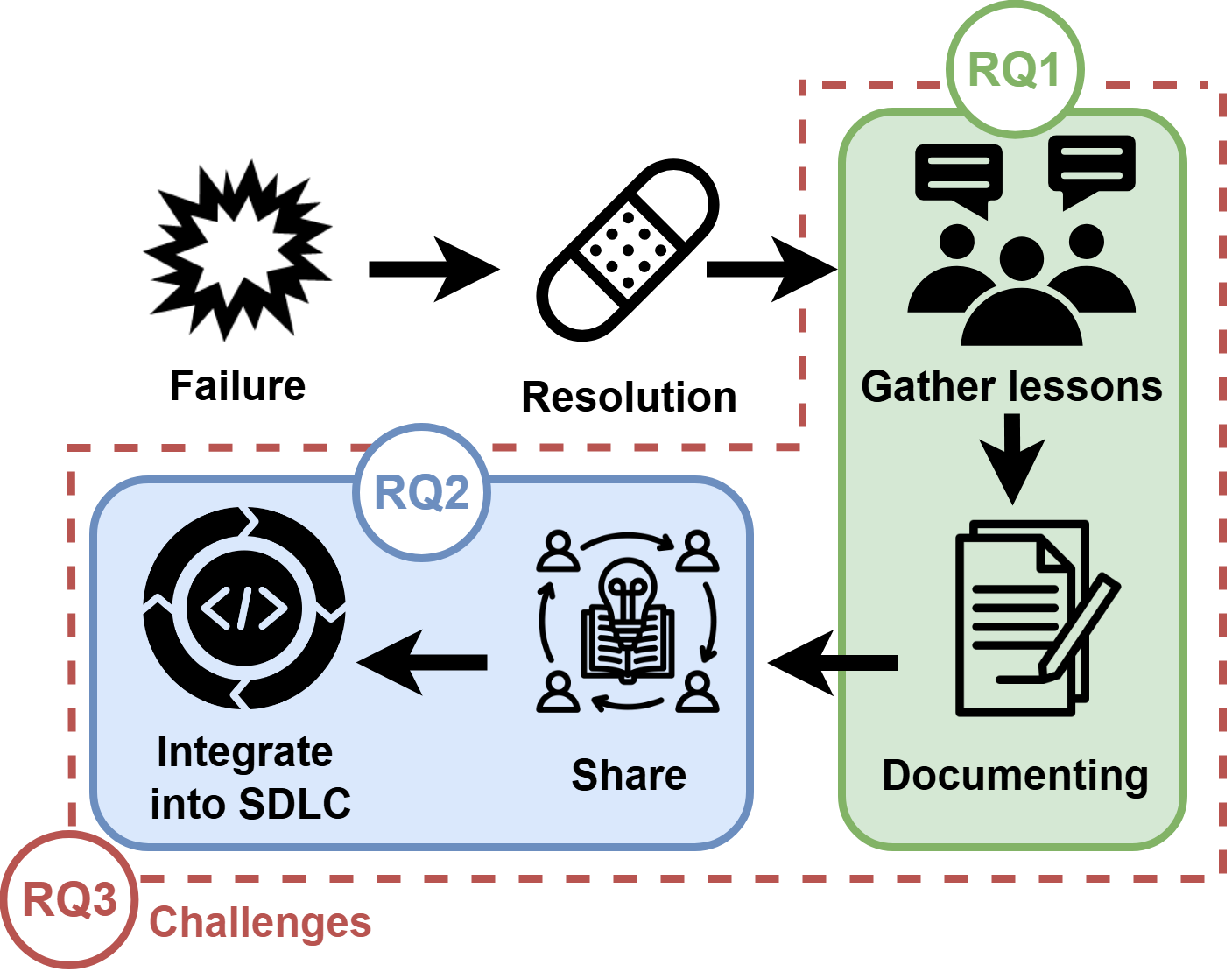}}
    \caption{
    The general process of learning from engineering failures (\eg vulnerabilities, incidents)~\cite{dingsoyrPostmortemReviewsPurpose2005, tiedemanPostmortemsmethodologyExperiences1990, collier1996defined, myllyaho2004review, stalhanePostMortemAssessment2003, dingsoyrAugmentingExperienceReports2001, kasiPostMortemParadox2008}.
    The dotted region indicates our study focus, annotated with the associated RQs. 
    }
    \label{fig:LearningFromFailure}
\end{figure}

Across all schools of software engineering thought, from ISO~\cite{internationalorganizationforstandardizationISOIECIEEE2018, ISO9001} to Agile~\cite{cockburnAgileSoftwareDevelopment2002, derby2006agile, kenschwaberScrumGuide2020} and beyond (standards~\cite{IEEEStandardSoftware2014}, frameworks~\cite{beck2000extreme, collier1996defined, fagan1977inspecting, fagan1999design}, and books~\cite{gilb1993software, beyer2016site}, etc.), guidelines agree that software engineers should also analyze failures and learn from them to improve their processes, as illustrated in~\cref{fig:LearningFromFailure}.
Software organizations employ postmortem methods~\cite{collier1996defined} to help teams reflect on their development experiences and problems to improve future development practices~\cite{dingsoyrPostmortemReviewsPurpose2005}. 
The goal of postmortems is to decrease the likelihood of future failures, by identifying lessons learned from reflecting on past failures and successes.

Existing literature on software postmortems can be broadly categorized into two areas: methods for conducting postmortems and experiences conducting them.
First, researchers have presented different methods to conduct postmortems~\cite{dingsoyrPostmortemReviewsPurpose2005, tiedemanPostmortemsmethodologyExperiences1990, collier1996defined, myllyaho2004review, stalhanePostMortemAssessment2003, dingsoyrAugmentingExperienceReports2001}.
These works have outlined approaches for identifying root causes, facilitating discussions, and gathering actionable insights.
Additionally, there is grey literature on postmortems, from
  software
    philosophies (\eg DevOps~\cite{kimDevOpsHandbookHow2021}),
    methodologies (\eg the agile principles~\cite{derby2006agile} and Scrum~\cite{kenschwaberScrumGuide2020}),
    standards (\eg ISO/IEC 90003~\cite{internationalorganizationforstandardizationISOIECIEEE2018}),
    and
    other guidelines (\eg Google's SRE handbook~\cite{beyer2016site} and the NASA Procedural Requirements~\cite{NASAProceduralRequirements}).
Second, researchers have also shared engineers’ experiences and the perceived challenges in conducting postmortems~\cite{desouzaExperiencesConductingProject2005c, dingsoyrOrganizationalLearningProject2007, kasiPostMortemParadox2008, washburnWhatWentRight2016, birkPostmortemNeverLeave2002, vieiraTechnicalManagerialDifficulties2019, gibekCaseStudyHow2024, sillitoLearningLessonsLearned2024}.
\added{These studies highlight issues such as time and resource constraints~\cite{desouzaExperiencesConductingProject2005c, kasiPostMortemParadox2008, vieiraTechnicalManagerialDifficulties2019}, difficulties in capturing and retaining lessons learned~\cite{lyytinenLearningFailureInformation1999, kasiPostMortemParadox2008, vieiraTechnicalManagerialDifficulties2019}, and cultural barriers that discourage open discussion of failures~\cite{desouzaExperiencesConductingProject2005c, kasiPostMortemParadox2008, vieiraTechnicalManagerialDifficulties2019}.}
However, neither body of work fully addresses the gap in understanding how engineers actually learn from failures in practice and the obstacles they face throughout this process. 


As a starting point, a few organizations have shared practices and experience reports \added{on learning from failures}.
\added{Practices are often structured around (1) gathering, (2) documenting, (3) sharing, and (4) applying lessons from failures~\cite{kasiPostMortemParadox2008}, as illustrated in~\cref{fig:LearningFromFailure}. Structured reviews or retrospectives are recommended to (1) \textit{gather} lessons from technical and socio-organizational factors that contributed to failures~\cite{dingsoyrPostmortemReviewsPurpose2005, beyer2016site, NASAProceduralRequirements}. For example,  Google describes a blameless postmortem approach to foster a culture of learning~\cite{beyer2016site}. Documentation practices are recommended to (2) \textit{document} not only root causes but also broader context, decision rationale, mitigation efforts, and lessons in an accessible and durable format~\cite{dingsoyrPostmortemReviewsPurpose2005, beyer2016site, NASAPublicLessons, NASAProceduralRequirements, johnsonSoftwareSupportIncident2000}. For example, organizations like NASA~\cite{NASAPublicLessons} and ESA~\cite{LessonsLearnedResources2024} maintain databases of lessons learned. (3) \textit{Sharing} often relies on both formal mechanisms like on-boarding training, cross-team reviews, and newsletters, as well as informal mechanisms like mentoring and reading clubs~\cite{dingsoyrPostmortemReviewsPurpose2005, beyer2016site}. However, literature on (4) \textit{applying} lessons learned is limited and their long-term effectiveness remains uncertain~\cite{popePostIncidentActionItems2024, sillitoLearningLessonsLearned2024, anandayuvarajIncorporatingFailureKnowledge2023}. A suggested practice is to create action items from lessons~\cite{dingsoyrPostmortemReviewsPurpose2005, beyer2016site}.}
 \deleted{However, recent progress in Large Language Models (LLMs) have enabled application of failure knowledge.  For example, Microsoft leverages automation to apply past failure knowledge to incidents, by helping understand~\cite{jinAssessSummarizeImprove2023}, conduct root cause analysis~\cite{zhangAutomatedRootCausing2024}, and suggest mitigation~\cite{ahmedRecommendingRootCauseMitigation2023a} for incidents.} 
\deleted{For example, Google describes a blameless postmortem approach, where incidents are documented without assigning blame, root causes are thoroughly analyzed, and actionable steps are identified to improve reliability and foster a culture of learning~\cite{beyer2016site}.}
\deleted{NASA follows a structured set of practices, such as convening review boards after incidents, maintaining a database of lessons learned~\cite{NASAPublicLessons}, and recommending systematic changes to improve future mission safety and accountability}


However, many organizations lack processes to study and learn from failures~\cite{vieiraTechnicalManagerialDifficulties2019,kasiPostMortemParadox2008, dingsoyrPostmortemReviewsPurpose2005, glassProjectRetrospectivesWhy2002, keeganQuantityQualityProjectBased2001}.
In addition, there is skepticism about how well these processes are followed in organizations that have them~\cite{kasiPostMortemParadox2008, sillitoFailingLearningStudy2024}.
The aforementioned exemplar organizations publish their policies, but rarely share engineers' experiences with them, \added{nor whether it leads to sustained learning, } especially beyond the postmortem room into the rest of the engineering process.
\added{To address this gap, we turn to HROs, which are theoretically expected to exemplify continuous learning and failure management~\cite{christiansonBecomingHighReliability2011}.}

\section{Research Questions}
In our analysis of the literature discussed in~\cref{sec:Background}, we observed that there is little empirical data on how software engineers learn from software failures\added{: how they gather, document, share, and apply lessons from failures}. 
We lack in-depth qualitative descriptions of industry experiences, creating a gap in our understanding of practices to learn from failures. 
We do not know how far practice is from theory, which practices are effective, nor the challenges in this engineering process.  
To fill this gap, we ask:
{\begin{RQList}

\item[\textit{RQ1}:] How do software teams gather lessons from failures?
\item[\textit{RQ2}:] How are these lessons shared \& integrated into the SDLC?
\item[\textit{RQ3}:] What are the challenges in learning from failures?

\end{RQList}}

\section{Methodology} \label{sec:Methodology}

\begin{figure}
    \includegraphics[width=0.98\columnwidth]{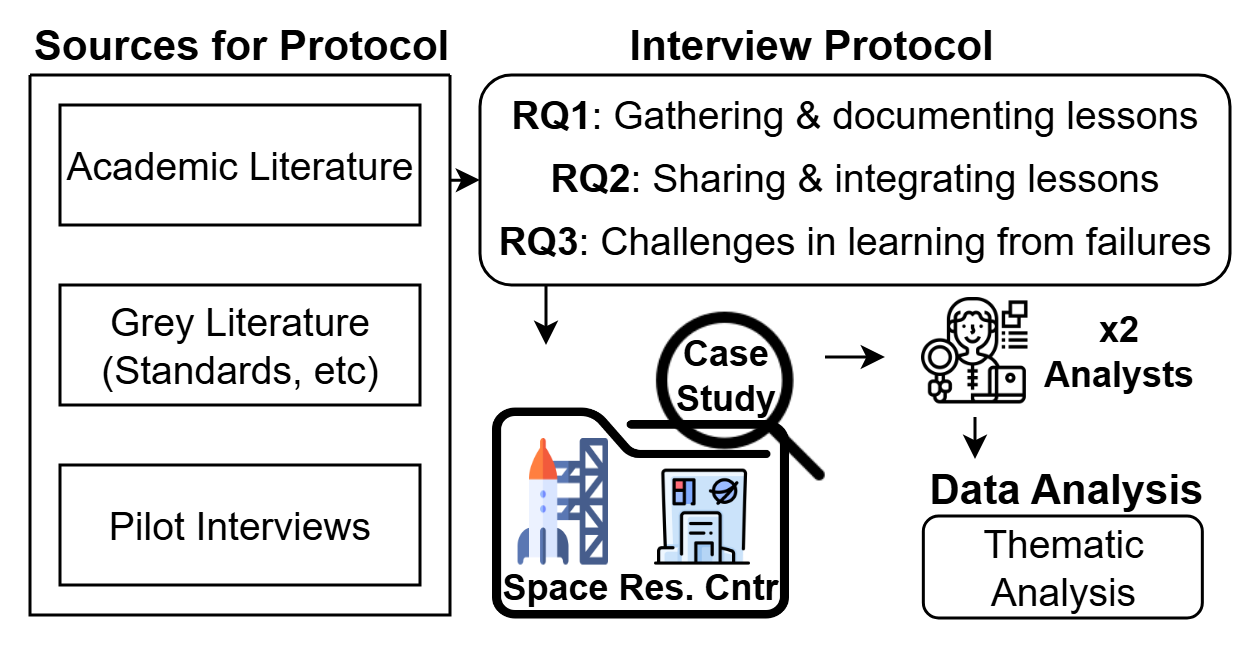}
    \caption{
    Study methodology.
    We situated our work as a case study at a space research center, complemented by pilot data from other HROs.
    Two author-analysts analyzed the data.
    }
    \vspace{-0.3cm}
    \label{fig: method}
\end{figure}

To answer our research questions, we conducted an exploratory case study at a national space research center through semi-structured interviews with research software engineers. 
We depict our methodology in~\cref{fig: method}.
To assess transferability of our findings, we also include data from our pilot study with software engineers from four other organizations. 
We describe
  our research design and rationale in~\cref{sec:Research Design and Rationale},
  the case study setting in ~\cref{sec:Case study setting},
  our interview protocol design and development in~\cref{sec:Protocol design and development},
  participant recruitment in~\cref{sec:Data collection},
  data analysis in~\cref{sec:Data analysis},
  and
  articulate threats to validity in~\cref{sec:Threats}. 

Our Institutional Review Board (IRB) approved this study.

\subsection{Research Design and Rationale} \label{sec:Research Design and Rationale}

Our research questions require insight into practitioner perspectives on the learning-from-failures process. 
In choosing a methodology, we considered two key factors. 
First, the phenomenon of interest is learning from failures in software engineering, which is a contextual process influenced by an organization's unique practices, culture, past experiences, and commitment to continuous improvement~\cite{leveson1995safeware,kalu2023reflecting}.
Second, the limited prior work in this area meant there was insufficient foundation for designing a closed-ended survey. 
Answering our research questions would therefore require in-depth, descriptive responses best obtained through open-ended discussions, opinions, and detailed observations. 

Given the preceding considerations, we therefore adopt a \textit{case study methodology} for our research design and analysis.
This approach enables us to investigate the phenomenon in depth in real-world context~\cite{DBLP:journals/corr/abs-2010-03525}.
For data collection, we took a \textit{semi-structured interview approach}~\cite{saldanaCodingManualQualitative2021}, allowing us to compare results across subjects while providing the flexibility to capture emergent insights~\cite{dejonckheere2019semistructured}.
The resulting data is anticipated to be exploratory and grounded in the perspectives and experiences of practitioners.

\subsection{Case Study Setting: A Space-related HRO} \label{sec:Case study setting}
Our study is set at the Department of Flight Software
at a major national space research center.
This department researches and develops flight software, ranging from embedded software to payload software for aircraft and spacecraft. 
The department is also the contact point for software quality assurance for safety-critical space systems.
The department currently supports around 15 space missions with its software expertise.


As a national space research center responsible for supporting mission-critical projects, it exemplifies characteristics of an HRO~\cite{christiansonBecomingHighReliability2011}. 
\deleted{This department operates in a high-risk context and shows sensitivity to operational failures.}
\deleted{Its structure and mission mandate a commitment to resilience and deferring to technical expertise.} 
The department’s role in software development for space missions, along with its responsibility for software quality assurance, aligns with the expectations of an HRO that must prioritize reliability, continuous learning, and failure management. 
Given these properties, we hypothesized that the organization would exhibit practices for learning from software failures, making it an appropriate site to conduct our study.

\added{The engineers in the department occupy a hybrid role: they are responsible for developing and maintaining mission-critical software, while also engaging in research activities such as experimenting with new programming languages and fault-tolerance techniques.
In a broad sense, they can be defined as Research Software Engineers (RSEs)~\cite{gothFoundationalCompetenciesResponsibilities2025, simsResearchSoftwareEngineering2021, segalWhenSoftwareEngineers2005}, who research and develop production-grade software, but are distinct from RSEs who primarily develop scientific analysis software.}

\subsection{Protocol Design and Development} \label{sec:Protocol design and development}
To develop the initial interview protocol, we studied the related literature (\cref{sec:Background}), which provided a  foundation of existing knowledge on learning from failures and highlighted gaps that informed our research questions and subsequently our interview questions.
We identified knowledge gaps and created the initial protocol, structured along typical postmortem and failure management processes.

We refined the protocol through practice interviews (2 subjects) and pilot interviews (5 subjects). 
For practice, the lead author interviewed two graduate students with prior SE work experience to iteratively improve the interview questions.
These practice interviews led to modifications across our interview protocol to improve depth and focus.
In the pilot, we conducted 5 interviews with practitioners from various HRO industries. 
\deleted{The subjects for the pilot study were recruited using convenience sampling, drawing from two groups: part-time professional master's students at our university (pilot interviews 1--3) and practitioners from the lead author's internship at another organization (pilot interviews 4--5).}
During these pilot studies we gathered insight about how these practitioners and their organizations learn from failures. 
We include data from the pilot study where feasible to evaluate transferability of our findings from the case study at the center to other organizations. 
The pilot study interviews led to modifications to improve focus on failure recurrence, communication, and automation.
In the Reflective Example section, new follow-up questions were added to explore whether issues reoccurred after initial fixes and what actions were taken to prevent recurrence. 
The Postmortem Analysis section was expanded to examine when issues are shared across teams and stakeholders. 
Additionally, a new Automation section was introduced to investigate desired tools for documenting and learning from failures. 
These changes aimed to capture deeper insights into long-term learning and prevention strategies.

\cref{tab:InterviewProtocolSummary} summarizes our final interview protocol.

{
\renewcommand{\arraystretch}{1.3}
\begin{table}
    \centering
    \caption{
     Summary of interview protocol.
     The full protocol is included in artifact. 
     The data from Topics E \& F are not analyzed in this paper.
   }
   \small
   \scriptsize
    \begin{tabular}{
    p{0.30\linewidth}p{0.6\linewidth}
    }
    \toprule
        \textbf{Topic (\# Questions)} & \textbf{Sample Questions} \\
    \toprule
         A. Reflective example (1) & Could you provide an example of a time when your team studied a software failure in order to gather lessons?  \\
         B. Postmortem analysis (4) & How does your team study and handle the failure? What is the process? \\
         C. Learning from failure (3) & How are the lessons learned from failures used for future development? How are they shared?\\
         D. Standardization (1) &  Are the practices discussed standardized across teams/departments/organization? \\
         \midrule
         E. Benefits and constraints (2) & Has your team benefited from practices to learn from failures? What are the constraints faced? \\
         F. Automation (4) & What type of automation would you like to help document (a) and learn (b) from failures? \\
    
    \bottomrule
    \end{tabular}
    \label{tab:InterviewProtocolSummary}
\end{table}
}
 
\subsection{Data Collection} \label{sec:Data collection}
\myparagraph{Target Population}
To effectively address our research questions, we focused on practitioners at an HRO, where learning from failures is essential to preventing recurrence and maintaining operational excellence.
We are interested in practitioners at all experience levels, from early-career  to expert practitioners. 
We gain valuable insight interviewing junior practitioners, who are still developing their ability to avoid repeating mistakes, as well as senior practitioners, who share lessons learned from past experiences.

\myparagraph{Recruitment}
\added{For the case study,} \deleted{As a summer intern} the lead author facilitated interviews within the organization \added{as a summer intern}.
The lead author's internship advisors, two of the secondary authors, provided a list of potential subjects for the study. 
These two authors were also interviewed for the study.\footnote{To eliminate bias, these authors-\textit{cum}-subjects did not participate in \added{data collection, coding, or} analysis, \added{except to provide feedback as member checking}.}
\replaced{A total of 10 interviews were conducted within the case study: 2 with these co-author subjects and 8 with other engineers.}{9 other candidates were invited to participate in the study, out of which 8 accepted.} 
\added{In addition, as stated in~\cref{sec:Protocol design and development}, we asses transferability of findings from the case study with the 5 pilot interviews.
The subjects for the pilot study were recruited using convenience sampling, drawing from two groups: part-time professional master's students at our university (pilot interviews 1--3) and practitioners from the lead author's internship at another organization (pilot interviews 4--5).}
All candidates were provided with a consent form which included brief questions on demographics (5 questions) and development practices (4 questions).

Demographic information for both pilot and case study subjects is summarized in~\cref{tab:Demographics}.

\myparagraph{Interviews}
All interviews were conducted \added{in English} by the lead author, via Zoom.
The median interview duration was 53 minutes.

{
\begin{table}[htbp]
    \centering
    \caption{
    Demographics of subjects in study.
    }
    \label{tab:Demographics}
    \scriptsize
    \begin{subtable}[c]{0.45\linewidth}
        \centering
        \caption{Demographics of pilot subjects.}
		\begin{tabular}{llcl}
            \toprule
           \textbf{ID} & 
           \textbf{Experience} &
           \textbf{Industry}
           \\
            \toprule
            PS1 & 1-5 years & Defense \\
            PS2 & 20+ years & Healthcare \\
            PS3 & 6-10 years & Manufacturing \\
            PS4 & 20+ & Automotive \\
           \multirow{5}{*}{PTeam} & \multicolumn{1}{c}{} & \multirow{5}{*}{Automotive} \\  
            \cmidrule(lr){2-2} 
            & 0 years & \\  
            & 6-10 years & \\  
            & 6-10 years & \\  
            & 6-10 years & \\  
            & 16-20 years & \\  
            \bottomrule
		\end{tabular}
        
    \end{subtable}%
    \hfill
    \begin{subtable}[c]{0.50\linewidth}
        \centering
        \caption{Demographics of case study subjects.}

		\begin{tabular}{llcl}
            \toprule
           \textbf{ID} & \textbf{Experience}\\
            \toprule
            S1 & 11-15 years \\
            S2 & 6-10 years \\
            S3 & 6-10 years \\
            S4 & 11-15 years  \\
            S5 & 1-5 years \\
            S6 & 6-10 years \\
            S7 & 20+ years \\
            S8 & 6-10 years \\
            S9 & 1-5 years \\
            S10 & 1-5 years \\
            \bottomrule
		\end{tabular}
        
    \end{subtable}
\end{table}
}

\subsection{Data Analysis} \label{sec:Data analysis}

Interview recordings were anonymized with classified information redacted, and  were auto transcribed by \url{www.rev.com}.
We conducted our analysis through a multi-stage coding process~\cite{campbell2013coding}, followed by thematic analysis~\cite{braun2006using}.
Two analysts (A1 and A2) analyzed the transcripts.
The lead author, Analyst A1, oversaw all stages of data collection, including protocol design and conducting interviews, and actively led every phase of the analysis.
A1 and A2 are graduate students with research experience in software engineering and software failure analysis.
The supervisors are scholars and practitioners in software engineering.

\myparagraph{Multi-stage coding}
\textit{Stage 1.}
We began with data familiarization~\cite{terry2017thematic}.

\textit{Stage 2.}
Next, we developed and refined our codebook.
To ensure reliability while optimizing resources, we applied the approach outlined by Campbell \etal ~\cite{campbell2013coding} and O'Connor \& Joffe~\cite{o2020intercoder}.
We selected a random set of 5 transcripts from the 10 case study interviews for the two analysts to code together (\ie multipy-coding).\footnote{While there is no universally agreed-upon proportion for this selection, O'Connor \& Joffe suggest that 10--25\% is typical~\cite{o2020intercoder}. 
To build our codebook, we randomly selected approximately 33\% (5 transcripts). 
This decision familiarized analyst A2 with our collected data and proposed memos, and brought consensus to our coding format.}
First, the two analysts memoed these transcripts independently. 
This process generated 343 memos, which were subsequently reviewed and coded by the two analysts through a series of meetings.
This process resulted in an initial codebook with 58 code categories.
Stages 3--5 discuss further codebook refinement. 

\textit{Stage 3.}
To evaluate the reliability of our coding, we integrated the methods proposed by Campbell \etal~\cite{campbell2013coding} and Maxam \& Davis~\cite{maxam2024interview}. 
Specifically, the two analysts independently coded a randomly selected transcript ~\cite{campbell2013coding, oconnor2019homesnitch}, and we assessed their level of agreement.
We used the percentage agreement metric, as recommended by Feng~\cite{feng2014intercoder}.
During this independent coding step, we achieved a 92\% agreement rate. 
All disagreements were resolved through discussion and iterative code refinement.
Campbell \etal \cite{campbell2013coding} suggest that this high level of agreement indicates a coding scheme reliable enough for use by a single analyst.  

\textit{Stage 4.}
Analyst A1 applied the refined codebook from Stage 3 to code the remaining 5 case study transcripts.
In parallel, Analyst A2 applied the codebook to the 5 pilot study transcripts. 

\myparagraph{Thematic Analysis}
After this process, we derived themes using the approach outlined by Braun \& Clarke~\cite{braun2006using}.
Through thematic analysis during our coding phase, we uncovered five distinct themes.
For brevity, this manuscript reports on the three themes that most effectively answer our research questions.

\begin{figure}[ht]
    \centering
    \includegraphics[width=0.95\linewidth]{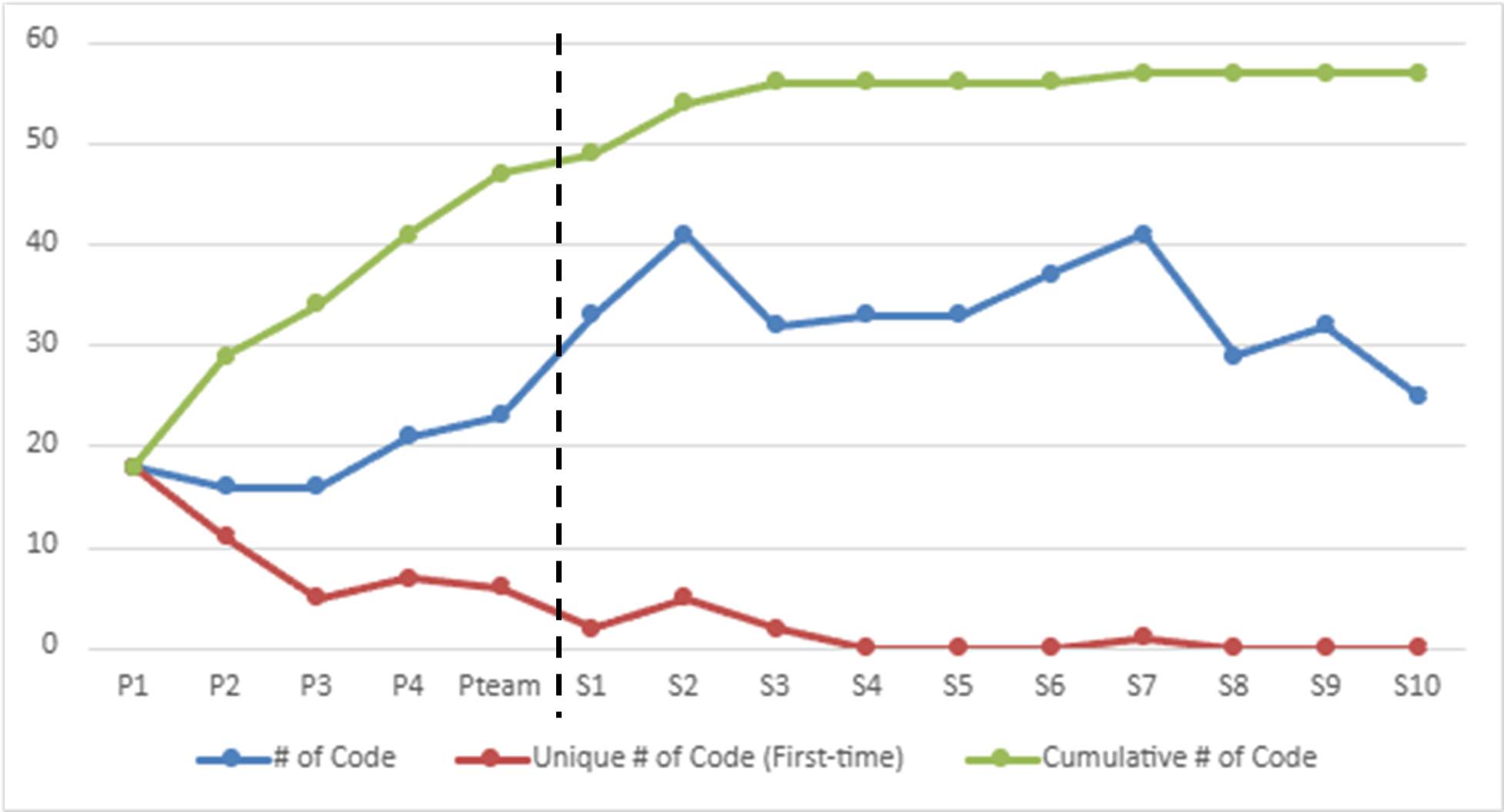}
    \caption{
        \small
        Saturation curve. 
        Interviews are plotted in the order in which they were conducted.
        Pilot interviews were conducted before the case study; they are separated by a dash-ed black line.
        Each interview covered many topics in detail (red line, blue line).
        However, as the green line shows, our results saturated--\emph{i.e.}, we stopped observing new perspectives--around interview~8.
    }
    \label{fig:saturation}
\end{figure}

To assess whether further sampling would reveal additional aspects of the phenomenon, we analyzed saturation across the 15 interview transcripts.
As recommended by Guest \etal~\cite{guestHowManyInterviews2006e}, we evaluated saturation by tracking the cumulative appearance of new codes in each interview transcript.
Each interview provided substantial data, with a median of 33 codes (\cref{fig:saturation}). 
Saturation was reached after the 8th interview (the 3rd case study interview), indicating that the case study provided new insights, and that its results are transferable to the pilot study.

\subsection{Threats to Validity} \label{sec:Threats}
We outline limitations in our study by discussing construct, internal, and external threats to validity~\cite{wohlinExperimentationSoftwareEngineering2012d} as well as author positionality.

\ul{Construct Threats} concern the alignment between our research methods and the concepts we intended to study. 
To mitigate such threats, we designed our interview protocol based on established literature on postmortems, failure analysis, and learning from failures in software engineering. 
Following established guidelines, we conducted both internal and external pilot interviews to ensure our our interview protocol would capture relevant data~\cite{chenail2011interviewing}.

\ul{Internal Threats} concern factors that may influence the validity of cause-and-effect interpretations.
As with any qualitative research, data interpretation carries an inherent risk of subjectivity. 
To enhance the reliability of our findings, we iteratively developed the codebook, involving a collaborative approach with multiple raters reviewing, discussing, and developing the codes, and achieved a high level of agreement.
We recognize the potential for bias since the case study was conducted at an organization where the authors were employed. 
\added{For member checking~\cite{birtMemberCheckingTool2016}, two senior subjects (who were co-authors but not involved in the analysis) reviewed our findings to validate our interpretations.}

\ul{External Threats} limit generalizability to other contexts.
We applied a case study methodology with semi-structured interviews.
This method explores a phenomenon in depth at one site, gaining detailed insight at the cost of less generalizability.
To provide some insight on the transferability of our findings, we supplemented our case study with interviews from subjects at four additional HROs. 

\ul{Author Positionality} refers to individual characteristics of the authors that may influence their analysis~\cite{holmes2020researcher}.
The lead author worked as a summer research fellow and conducted interviews at the organization. 
Two other authors are employed at the center.
A third author works at a comparable organization in another country.
These professional positionalities afford strengths and weaknesses.
Employment relationships and experience in similar working environments permit data collection and improved interpretation of contextual and technical aspects of the data.
However, they also add biases to reflect an employer or a class of organization in a positive light.
To mitigate that bias, all data analysis involved a second author-analyst with no relationship to these organizations.

\section{Results and Analysis} \label{sec:Results}

We organize our results into three themes that answer our research questions.
A summary of results is outlined in~\cref{tab:ResultsSummary}.
\added{Our analysis reveals 6 new findings and 5 findings that have been partially discussed in prior works.}

\begin{table*}[ht]
\centering
\small
\renewcommand{\arraystretch}{1.1}
\caption{
Findings from our thematic analysis mapped to research questions. \added{The \textit{Lit.} column notes prior works discussing related findings. The \textit{Pilot Sub.} column indicates pilot subjects whose interviews support transferability of the finding.}
}
\begin{tabular}{p{0.015\linewidth} p{0.76\linewidth} p{0.09\linewidth} p{0.067\linewidth}}
\toprule
\textbf{RQ} & \textbf{Section and Summary} & \textbf{Lit.} & \textbf{Pilot Sub.}\\
\midrule

RQ1 & (\cref{sec:Theme 1.1}) Lesson gathering is ad hoc, informal, \& individually driven; reflection is not routine. & \hspace{-1pt}~\cite{kasiPostMortemParadox2008, vieiraTechnicalManagerialDifficulties2019} & 1, 2, 3, T\\
 & (\cref{sec:Theme 1.2}) Documentation practices are inconsistent, leaving failure knowledge fragmented \& largely tacit. & \hspace{-1pt}~\cite{kasiPostMortemParadox2008, vieiraTechnicalManagerialDifficulties2019} & 1, 2, 3, T\\

RQ2 & (\cref{sec:Theme 1.3}) Lessons are shared via informal conversations \& senior mentorship within teams. Cross-project sharing is rare. & \hspace{-1pt}~\cite{kasiPostMortemParadox2008, lyytinenLearningFailureInformation1999} & 1, 2, 3, 4\\
 & (\cref{sec:Theme 1.3}) Applying past lessons is driven by individual memory or initiative. Improvements aren't part of structured cycle. & \textemdash{} & 1, 2, 4, T\\
 & (\cref{sec:Theme 2.3}) Some examples of \textit{ad hoc} learning exist, but they are exceptions \& not integrated into broader practice. & \hspace{-1pt}~\cite{sillitoFailingLearningStudy2024} & 1, 2, 4, T\\

RQ3 & (\cref{sec:Theme 2.1}) Recurring failures are common due to lack of structured processes for reflection \& improvement. & \textemdash{} & 1, 3, 4, T\\
 & (\cref{sec:Theme 2.2}) Teams often assume good practices suffice to prevent failures: masks systemic gaps \& limits formal learning. & \textemdash{} & 1, 2, 4\\
 & (\cref{sec:Theme 3.1}) Time \& resource constraints limit opportunities for failure learning; structured methods are seen as burdensome. & \hspace{-1pt}\rlap{{\footnotesize\cite{kasiPostMortemParadox2008, desouzaExperiencesConductingProject2005c, vieiraTechnicalManagerialDifficulties2019,lyytinenLearningFailureInformation1999, dingsoyrPostmortemReviewsPurpose2005}}} & 1, 2, 3, 4, T\\
 & (\cref{sec:Theme 3.2}) Informal, research-oriented SDLC practices reduce emphasis on formal learning from failure. & \textemdash{} & \textemdash{}\\
 & (\cref{sec:Theme 3.3}) Knowledge is often lost due to team turnover \& poor documentation of lessons learned. & \textemdash{} & 1, T\\
 & (\cref{sec:Theme 3.4}) Failure knowledge is fragmented across platforms, hindering accessibility \& application. & \textemdash{} & 1, 2, 4, T\\

\bottomrule
\end{tabular}
\label{tab:ResultsSummary}
\end{table*}

\subsection{Theme 1: Failure Learning is \textit{Ad--hoc}} \label{sec:Theme 1}
\added{We report findings on how engineers gather, document, share, and apply lessons from failures.}

\subsubsection{Gathering lessons from failures is informal} \label{sec:Theme 1.1}

Across all subjects, there was no structured or consistent process for gathering lessons learned from failures; instead, such efforts were \textit{ad hoc}, informal, and dependent on personal initiative.
All subjects stated that formal processes for reflection were absent. 
As \SubjectFive noted, \myinlinequote{I don't think that we have a formal process on how to make sure that [failures] doesn't happen again.}
Others echoed this absence, with \SubjectSix stating, \myinlinequote{We don't have a very strict process...not really,} and \SubjectThree explaining that group reflection is not part of their routine: \myinlinequote{There's not a big reflective period where we say, `Okay, this error happened...how should we do it in the future?'}
\SubjectOne emphasized \myinlinequote{we study the problem in order to fix the problem and not...to retrieve some knowledge out of it.}
Instead, participants described informal and improvised efforts to reflect on failures. 
These were typically team-driven and irregular, such as a one-time feedback meeting described by \SubjectEight: \myinlinequote{We're going to do this actually in a couple of weeks...but it's not standard.} 
\SubjectNine confirms, \myinlinequote{We do it...on the fly...But negative point is of course that there's no real structure, it's more like sharing experiences.}

When it does occur, reflective practices are often initiated by individuals with personal motivation, rather than organizational norms. 
\SubjectFour explained, \myinlinequote{I try to implement those kinds of good practices...documenting the issues, learning from them, to reflections...but not as part of the main software development process.} 
\SubjectFive emphasized that improvements often arise because someone personally pushes for them: \myinlinequote{Steps...are not in a process defined by itself, but more on the personal motivation of some developers.}
Participants also suggested that issues perceived as more critical, architectural, or linked to external reviews were more likely to trigger reflection. 
As \SubjectTwo observed, \myinlinequote{Issues are studied and lessons learned gathered based on criticality [determined by potential] damage.} 
Notably, such efforts often occurred at the discretion of senior members. 
As \SubjectTwo noted, \myinlinequote{Lessons learned...it’s mainly up to...the seniors on the group.}
While some senior members, like \SubjectSeven, took initiative to reflect on what went well or poorly, these efforts were neither widespread nor institutionalized: \myinlinequote{Normally for me it’s always necessary after a project to checking what’s good, what was wrong, what would be better and so on...[but] this is not reviewed by this department.}

Notably, a few subjects described practices that revealed that they were confused about what constitutes a lesson learned, equating fixing bugs, documenting issues, or general discussions with reflective learning. 
As \SubjectSix explained, \myinlinequote{There was this issue and the lesson learned will be...to fix this issue.}
\SubjectTen also incorrectly interpreted postmortems, \myinlinequote{we have a scrum meeting and we speak about the things...finished...the plans we have for the future...so that maybe counts as postmortem.}

\subsubsection{Documentation of lessons learned is inconsistent \& fragmented} \label{sec:Theme 1.2}
Most subjects (8/10) reported that documenting lessons learned was \textit{ad hoc}, informal, and left to individual discretion.
\SubjectFour explained, \myinlinequote{when we close the issue, it is not mandatory to document what did you do...[it's a] personal decision. We don't have that process...I know from experience that this is important so I do it.} 
Half of the subjects indicated that lessons learned are not documented, as \SubjectSix admitted, \myinlinequote{I don't think they are documented, pretty much at all.} 
Most subjects shared that they, \myinlinequote{track issues, but not for lessons.}
While some participants, like \SubjectThree, acknowledged that documentation might happen occasionally, they emphasized the lack of a defined process: \myinlinequote{Yes, it might be, but it's not like there's a process for that.} 
As \SubjectTen suggested, \myinlinequote{Firstly maybe you have to standardize what you should write.} 

In absence of formal processes, subjects rely on tacit knowledge rather than accessible records.
When asked about lessons learned, \SubjectTen stated, \myinlinequote{Not really lessons learned, no. I think they're more internal...we learned from it.} 
Likewise, \SubjectEight emphasized the implicit nature of such knowledge: \myinlinequote{Not explicit, but implicit...you learn how to do it a better way, but this is just internal knowledge.} 
\SubjectTwo, \SubjectSix, and \SubjectTen shared examples where even when recurring issues are discussed, the knowledge is rarely written down, \myinlinequote{I don't think we document it anywhere to be honest...Maybe it would be a good idea.}

Where documentation does occur, subjects described various methods used to capture failure knowledge, highlighting that it is distributed across multiple platforms.
\SubjectFive explained that issue severity affects where documentation is placed: \myinlinequote{If it's...bigger issues...then it might get its own GitLab issue...If it's just a local discussion...then I don't think it gets documented.} 
While GitLab was frequently mentioned as the primary repository for failure knowledge, many participants also referred to inconsistent practices of storing knowledge in supplementary tools.
Other participants described failure documentation being scattered across \myinlinequote{merge requests} (\SubjectFour), \myinlinequote{meeting notes} (\SubjectTwo), and \myinlinequote{scientific documents...showing the lessons learned} (\SubjectSeven).
The most senior participant, \SubjectSeven, also reported contributing insights to a software engineering network wiki within the organization, \myinlinequote{We have the software engineering network and several lessons learned inside...coming from...my experiences.}
Although subjects mentioned wikis, they did not use or refer to them, showing how even widely available resources like wikis remain underutilized.
\SubjectTwo explained, \myinlinequote{When you look to the wiki page called ‘lesson learned from project,' it's a few, not every project has one.}
Most participants were not well aware of these wikis, suggesting that such contributions may not be widely known or accessed.
This fragmentation of knowledge is further discussed in~\cref{sec:Theme 3.4}, where it poses barriers to application of knowledge and team coordination.

\subsubsection{Lessons learned are shared \& applied informally} \label{sec:Theme 1.3}
All case study participants reported that failure knowledge is shared through informal discussions rather than structured processes within teams.
As \SubjectFour explained, \myinlinequote{Not officially, but it is common that during the day I discuss something with my team members...but not in an official way.} 
\SubjectNine echoed \myinlinequote{[It's] just on the go.} 
Most subjects mentioned stand-up meetings as a space to share blockers and informally learn from past issues:
\SubjectThree stated, \myinlinequote{[In] stand-up meetings... [we] discuss...obstacles...[and] if someone has had the problem before [they can share]...but it's not in every project.}
Some subject also shared more casual discussion, as \SubjectTwo noted,
\myinlinequote{Sometimes in coffee break...we had this issue and we tried to learn this...in short discussion...I learned from their lesson learned, then I personally will keep it in mind...when I face a similar problem.} 
While these informal discussions sometimes result in useful insights, they are not systematically shared. 

Subjects shared that lessons often remains siloed within teams, as cross-project sharing is circumstantial. 
\SubjectOne admitted, \myinlinequote{Very rarely we report in a bigger round...that we experienced a failure...So it completely stays in the team and will not be exchanged.}
Others shared that external sharing depends on severity or criticality. 
As \SubjectTwo explained, \myinlinequote{If it's relevant for other projects or a very critical situation, then this information is spread.}  
While some reported knowledge carried over through individual mobility across projects, as noted by \SubjectThree:\myinlinequote{[If] I work in one project, get some problems, start working on another project, and of course I bring this experience and try to communicate this,} few described structured mechanisms to facilitate that transfer.

Subjects highlighted that the most consistent form of failure knowledge transfer occurred through informal mentorship from senior developers, who act as repositories of failure knowledge.
All subjects cited \SubjectSeven (the most experienced) as their go-to person for resolving issues. 
\SubjectNine emphasized this reliance: \myinlinequote{I...have to ask [\SubjectSeven]...[using] his experience and we can then decide how we want to avoid problems or...to improve.}
\SubjectSix echoed, \myinlinequote{[When] I have issue...I...just go to [\SubjectSeven's] office and ask...how would you tackle it?...Normally it's knocking the door and asking, `Hey, do you have experience on [this]?'}
While this method was appreciated by junior developers, 
others noted its limitations. 
\SubjectOne highlighted that this one-person dependency led to inefficiencies resolving recurring failures:

\blockquote{A couple [of] situations...where issues...were recurring...[and often] someone else on the team...[would recollect], `Oh yeah, I remember I faced this too. You have to do...' So that's probably the worst case...to handle those things, but that's how it's done...You have to hope...[to find] someone [who remembers that they] already experienced it...there's not a process to look it up.}

The absence of a centralized source of knowledge meant that even recurring issues had to be rediscovered unless someone remembered them. 
As \SubjectOne put it, \myinlinequote{This could happen only if the person who fixed it...is still there. I never experienced someone say, ‘We had this issue before and fixed it like this.’}

Application of lessons learned was \textit{ad hoc}, driven by individual initiative, and typically occurred only when subjects (5/10) recalled a similar issue or knew where to find relevant documentation.
As \SubjectThree described, \myinlinequote{I...couple of times...searched through the issues...to figure out...what I did last time...but it's not in the process...to prevent...these issues again.} 
\SubjectTen does the same, \myinlinequote{[but] I don't know if other people do it.}
\SubjectFour recounted using past issues for \myinlinequote{training} and \myinlinequote{to guide users to design their applications,} but acknowledge that it is a personal practice.
Similarly, \SubjectNine says, \myinlinequote{sometimes lesson learned...are referred when someone new joins...[but] it's not too often...last time...was two years ago.}
\SubjectFive acknowledged that even with recurring issues, their issue notes are not referenced: \myinlinequote{Issues appear over and over again...[however] issue...meeting notes are not really checked. I’ve...maybe twice or thrice.}
Others expressed that referencing past documentation is not common practice: \SubjectEight remarked, \myinlinequote{No, I think it's not so.}

\deleted{To enable feedback cycles, \SubjectSeven mentions questionnaires, though no other subject was aware of them, and they remain unused.}
To enable feedback cycles, \SubjectSeven mentioned questionnaires:
  one to reflect on failures, \myinlinequote{a questionnaire to discuss...what's wrong?...can this happen again?},
  and another to apply lessons during, \myinlinequote{design and getting...requirements...to get the criticality of system.}
However, when we pointed out that no other subject referenced these, \SubjectSeven admitted,
\myinlinequote{Normally you should use it always...But people don't...we cannot [enforce] everyone to do it,}
and as a result, \myinlinequote{failures are made again and again.}
As \SubjectTwo said, \myinlinequote{You'll [have to] be self-motivated to search the wiki...GitLab, ask the seniors...otherwise...there's no process to bring lessons down to you.}

\subsubsection{Transferability} \label{sec:Transfer - Theme 1}

The pilot interviews matched the case study data: gathering, documenting, sharing, and applying lessons from failures were \textit{ad-hoc}.

\added{Pilot subjects shared that failure knowledge was gathered informally and often depended on personal motivation (cf.~\cref{sec:Theme 1.1}).}
For example, the \PilotTeam described a one-off, \myinlinequote{brainstorming workshop...where we discussed...production issues we have being observing...and not having those repeat again,} but noted this was an exception, not the norm.
\PilotSubjectThree admitted, \myinlinequote{I think I was the only one who would [reflect].} 
\added{Subjects stated the absence of formal reflection practices, with 
\PilotSubjectOne acknowledging, \myinlinequote{we don't have any kind of formal investigation process. It's more, `just go fix it.'}}

\added{Documentation practices of failure knowledge in the pilot study mirrored the case study’s absence of structured practices, due to time and resource constraints (cf.~\cref{sec:Theme 1.2}). 
\PilotSubjectOne explained \myinlinequote{we're very short-staffed...we don't have time to write documentation.} 
\PilotSubjectTwo noted that postmortems are simply recorded as \myinlinequote{meeting minutes} without structure, since \myinlinequote{there isn't a specific template for the postmortem data,} and as a result, \myinlinequote{I don't believe we are at a point where we are collecting all that [postmortem] data.}}

\added{Application of lessons from failures was limited, and knowledge transfer often relied on informal mentorship (cf.~\cref{sec:Theme 1.3}).
\PilotSubjectOne reflected, \myinlinequote{I've never heard anyone say, `how can we make sure [failures] don't happen again?'}
\PilotSubjectTwo stated that failure knowledge was \myinlinequote{not referenced once [the project] is closed.}
\PilotSubjectFour noted that, \myinlinequote{action items [after failures] are put in a [wiki] or an email, so someone would manually track it...but I don't know if anybody follows up and ensures that the action items get done.}
\PilotSubjectOne explained that knowledge transfer was passive and relied on junior engineers  taking initiative: \myinlinequote{They don't really incentivize...staff engineers to mentor [juniors]...to give them knowledge...it's more on the younger engineers to ask for it.}
\SubjectFour similarly remarked that lessons are \myinlinequote{out there, but it's not specifically addressed in training,} with the \PilotTeam adding \myinlinequote{they just watch how other team members are responding.}}

\subsection{Theme 2: Recurring Failures and the Illusion of Effective Failure Management} \label{sec:Theme 2}
\added{Although engineers often believed they manage failures effectively, recurring incidents reveal gaps that undermine this perception.}

\subsubsection{Recurring failures persist due to lack of structured learning} \label{sec:Theme 2.1}

All subjects reported facing the same types of issues multiple times, \eg across different projects or development stages. 
\SubjectSeven, the most senior participant, noted, \myinlinequote{failures are made again and again...the same problems.} 
\SubjectTwo says that it happens across projects:
\myinlinequote{Same problems but different places because it's got a different project, different requirement.}
\SubjectSix sees it across developmental phases, \myinlinequote{we're...committing the same mistakes over and over just in different phases.} 
\SubjectNine complains that rather than preventing recurrence during development they fix it during maintenance:
\myinlinequote{We...face the same issues again instead of implementing it in a proper way.}

Most subjects recognized that recurring failures stem from a lack of structure in learning from failures.
\SubjectOne described, \myinlinequote{Couple situations...in the last couple of years where issues in our software...were recurring...again and again, but there wasn't a process to handle those...So that's probably the worst case, how to handle those, but that's how it was done actually.} 
Subjects allude recurrences occurred because lessons were not formally recorded, generalized, or communicated.
\SubjectOne added, \myinlinequote{after a while someone forgot...and this issue popped up...again...we discussed this already five months ago.}
\SubjectTwo recounts another example, \myinlinequote{we found an issue...we fix it...and then we got this problem again...[because] fix [is not] discuss[ed] on the group level...others...were not aware about the issue at all.}
\SubjectTwo alludes to a deeper underlying issue, where problems are solved on a case-by-case basis rather without gathering generalized lessons:
\myinlinequote{The problem...[of] problems happening again and again because...we solve it by scenario based and every time we got to actually to solve it again...and we need to provide a generic solution, not scenario based solution.}
Without structure practices to gather and communicate lessons, local fixes remain siloed, and failures reoccur.

\subsubsection{Confidence in general practices undermines the perceived need for learning from failures} \label{sec:Theme 2.2}

Many participants believed that adherence to such general software engineering practices, was sufficient to prevent the recurrence of failures, reducing the perceived need for structured failure learning processes. 
\SubjectNine explained, \myinlinequote{we have a structure there already...we have our requirements...a test-driven approach...guidelines, standards.} 
\SubjectThree described relying on, \myinlinequote{coding guidelines...defensive programming...templates...unit tests...and peer review before merging...to avoid...errors...but it's not that I sift through [past] issues.} 
\SubjectSix emphasized frameworks were designed to prevent known issues: \myinlinequote{we tried to come up with these reusable frameworks...because history repeats itself.}
A few subjects, like \SubjectFour, felt that documenting bug reports sufficed, \myinlinequote{issue reappearing... [is] prevented by...adding example...bugs...for the issue.}
\SubjectFour emphasized that general measures embed years of industry experience, and is sufficient to prevent recurring failures:

\blockquote{We use several general measures to avoid these...[recurring failures]...we rely on best practices, guides, standards...because those collect years of experience [and reflection] in software development from industry...That’s what prevents...issues.}

However, some participants acknowledged limitations in relying solely on best practices. 
\SubjectOne \added{noted the overhead of following best practices from the organization's software guideline}, \myinlinequote{[It's] a 150--200 page PDF with a lot of rules...even I can't tell you all the rules. [We] should know where this is, but not learned by heart...[it's] not [referenced] that often.}
Consequently, although the guideline requires them to gather lessons learned, they do not seem to do it consistently, as \SubjectTwo shares, \myinlinequote{The standard from the...Space Agency...[requires us to] collect lesson learned...but this doesn't apply for all of our projects.}
\SubjectSeven criticized the superficiality of such guidance \added{in describing how such processes should be implemented in practice}: \myinlinequote{With the standards...the common agreement is...so weak...It didn't help you normally...[because] they don’t share too much knowledge,} \added{consistent with prior findings~\cite{chenContentsUtilityIoT2024}.}

Notably, some subjects attributed recurring failures to the inexperience of junior engineers -- masking systemic gaps in how failure knowledge is transferred.
\SubjectSeven observed, \myinlinequote{I see [failures] happened again and again because...young software engineers...didn't have the experience and the sharing of experiences is limited.} 
They defended their on-boarding process, claiming it mitigates these gaps:
\myinlinequote{There aren't really gaps, there's training. The only issue we have is time to get the lessons and [applying] the lessons. That's why we have an on‑boarding process.}
Responsibility for avoiding recurring failures was shifted to individuals, particularly newcomers, rather than organizational processes to learn from failures. 
\SubjectSeven explained, \myinlinequote{as long as the process is understandable for new employees...for experienced engineers it might not be necessary...[they already have] a big knowledge base.} 
This assumption risks perpetuating knowledge silos and obscuring structural challenges in failure knowledge transfer.

\subsubsection{Ad hoc learning sometimes prevents recurrence--but is rare} \label{sec:Theme 2.3}

In a few cases, serious or recurring failures prompted technical or process improvements, though these were typically driven by impactful incidents and individual initiative rather than structured processes.
\SubjectSeven described a recurring, 

\blockquote{big fail[ure]. The project manager called and I hear...crying.
[He says:] `we cannot start the satellite...nothing works.' \newline
[I reply:] `Yeah, I know this failure, it was fixed three months ago'. \newline
[From this failure, our] lessons learned was...this modeling group...a test process...[and] a research project.}

In another case, \SubjectSix explained how a post-mission reflection led to improvements: \myinlinequote{From results of mission one...we learn a lot... [and] we are applying it for the new mission.}
In a different case, \SubjectTwo recalled that after a recurring issue, the team learned, documented a guideline, and shared it: \myinlinequote{we...define a guideline...how to...and what not to do...to not get this problem,} and as a result the problem did not recur \myinlinequote{because the people who joined afterwards followed the same instructions.}
\SubjectThree noted that from a prior failure, they, \myinlinequote{have a lot of checks beforehand...so later you don't have as many pitfalls.}

\subsubsection{Transferability} \label{sec:Transfer - Theme 2}

\added{Pilot subjects also reported facing recurring issues due to lack of structured practices to learn from failures. 
\PilotSubjectFour admitted, \myinlinequote{we have multiple root cause analyses with same [root cause], but we never prioritize getting it fixed...[so] we come across the same issues over and over...it gets worse and worse.}
\PilotSubjectThree acknowledges the lack of emphasis to learn from failures, \myinlinequote{they don't care about failures. They just work past it and continue making failures.}}

\subsection{Theme 3: Challenges within the Learning-from-failures Process} \label{sec:Theme 3}
Themes 1 and 2 highlighted the ad--hoc nature of learning from failures.
Since many subjects were aware of the ill effects, we examine why more robust processes have not taken root.

\subsubsection{Lack of time and resources} \label{sec:Theme 3.1}
All participants emphasized that limited time and competing responsibilities constrain their ability to engage in structured failure learning. 
Subjects describe wearing multiple hats at the center, as \SubjectTwo listed, \myinlinequote{we are not solely software developers, we're also researchers, sometimes project managers, systems engineers.} 
Each of them have various responsibilities as \SubjectOne outlined, \myinlinequote{our job is not [just] the development, you have project meeting...paper stuff...usually more than one project...bureaucracy...some other additional tasks...deadlines.}
\SubjectFive observed, \myinlinequote{most employees are...drowned in project work.}
As a result, time for such reflective processes is limited, as \SubjectOne questions, \myinlinequote{I don't know if we would have time to designate to such a process.} 
Because, as \SubjectFour explained, \myinlinequote{the main issue is time, because...you cannot...spend much time thinking about...issues.} 
All subjects echoed this, including \SubjectTwo: \myinlinequote{sitting and reading or creating new incidents...will take a long time.}

\SubjectThree and \SubjectSeven pointed out budget constraints:
\myinlinequote{You have only funding for project, and normally you [would be] doing  lessons learned afterward.}
As a result, \SubjectFive says, \myinlinequote{I don’t know if people would prioritize learn[ing].}
\SubjectNine reflected on this trade-off, \myinlinequote{it's a matter of how much time do you invest, creating a structure and how much use you have out of it.}
\SubjectTen says that the team must believe in the trade-off, 
\myinlinequote{I think the team has to see postmortems as sensible and worth the time.}

Concerns were also raised that formalized reflection meetings could become bureaucratic or burdensome. 
\SubjectThree noted, \myinlinequote{we have all the bureaucracy, every additional meeting puts a big hole in your calendar...the motivation to attend meetings decreases...if you make another meeting...[to] talk about reflection, then everyone will hate you...it will end up a big bureaucratic process where everyone just pukes again...[where] the process is bigger and more complex.} 
\SubjectEight echoed this sentiment, stating, \myinlinequote{if it’s something like twice a month or even once a month, for me it will become a little bit tedious.} 

While discussing challenges, however, many subjects acknowledged that learning from failures will save time and is worth the investment.
\SubjectOne admits, \myinlinequote{That's a bad excuse not having the time when we save time by this.}
\SubjectFive echoes, \myinlinequote{maybe that's even more of a reason to take the time to do that. So...I would like to try.}
Furthermore, \SubjectFive admits that it will save time by preventing recurring issues: \myinlinequote{It would be helpful to have a process to learn from mistakes...because it takes a lot of time if people make the same mistakes over and over again.}
\SubjectNine suggests balancing with existing process:
\myinlinequote{if you do it [with] milestones...then it might be a huge benefit.}
\SubjectSix proposes periodic reflection sessions:
\myinlinequote{I'm in favor of...time allocated to that. I will...propose to have these kind of discussions...I think...it comes at the cost of time and personal resources [but] at the end I think everybody benefits from it. So actually the loss not a loss, but a gain.}
\SubjectFive unwittingly admits that postmortems might work better than current ad-hoc one-on-one learning, \myinlinequote{[it] would make more sense if you do it with...whole team and not just with one person.}

\subsubsection{Informal \& flexible SDLC limits learning structures} \label{sec:Theme 3.2}
Participants frequently described their SDLC as informal, flexible, and research driven, distinct from traditional or agile industry models. 
\SubjectSix stated, \myinlinequote{we're mostly a research institute...we don't have standardized processes...it is quite informal.} 
When asked if they reflect on failures, \SubjectFour summarized:
\myinlinequote{We are not doing that...because we are not totally following agile...We use some aspect of agile, but...[in] space project you don't see total agile.}
Many subjects equated reflective practices with agile development, and contrasted their development with agile and tech companies, as \SubjectOne further explained:
\myinlinequote{probably big [tech] companies...[have] a more agile mind than us because here you have old processes because space is not agile and it's not wise to do it.}
As a result subjects are apprehensive about reflective processes, as \SubjectFive says, \myinlinequote{I feel like it's not really feasible for the way we work}.
While institutional software guidelines may state to follow a failure process, \SubjectOne explained they are not actively enforced: \myinlinequote{There are guidelines from the [organization] and...they state that issues need to be documented...but they are not enforced.}

Many acknowledged that other engineering disciplines within the center, particularly hardware teams, follow more rigorous post-failure processes due to the high cost of mistakes. 
As \SubjectFive observed, \myinlinequote{If you do something stupid...in software then you just update your software...but in hardware, you burn money...because it's harder to fix in the end.} 
\SubjectFour added: \myinlinequote{So...they have to move more careful than software people do},
as did \SubjectOne, \myinlinequote{And I think for this reason there is probably a process.}
As a result, \SubjectOne notes, \myinlinequote{In other institutes...doing satellite hardware, they have [a] knowledge base on what are typical faults or failures...in our institute there's nothing like this.}
This believe that software failures are less expensive may reduce pressure to institutionalize learning.
\SubjectSix contradicts the other subject by noting the high cost of software fixes after deployment~\cite{zage2003analysis,McConnell2001once,stecklein2004error},
\myinlinequote{even with...a software update...[on] something flying on the [space station], you need to...[do] software approval and this can take...half a year to a year...[Even if] a software update..[is] `free'...it's not easy and you want to get the software right the first time.}

Despite the lack of formal learning structures, the research aspects of the subjects' role offered unplanned reflection. 
Per \SubjectFour:

\blockquote{We do...project development...also research. For research...when you are doing that process...you usually discover some pitfalls or issues...when you are doing a publication...I think that's an opportunity to think and to improve.}

\subsubsection{Knowledge loss from team turnover} \label{sec:Theme 3.3}
Most participants (6/10) noted that developers frequently join and leave teams, resulting in loss of failure knowledge. 
\SubjectTwo exclaims, \myinlinequote{how will you train a team that's changing every six months?}
Subjects expressed that relying on tacit or interpersonal knowledge transfer is insufficient under such turnover. 
\SubjectTwo admitted to knowledge loss from current ad-hoc one-on-one knowledge transfer, \myinlinequote{Even if [seniors] bring [failure] knowledge to developers...every six months someone will join and someone will leave.}
\SubjectNine shared her experience, \myinlinequote{I had a senior and after I joined he left. So I was alone with no experience and it's hard [because] so much knowledge was just gone within seconds.} 
Many subjects, like \SubjectTwo, felt that this motivated more documentation:

\blockquote{[Lessons] should be explicitly documented because experience is just related to one person...losing this person is losing the knowledge, losing the lesson learned...because there's no process to say `part of your project is to write lesson learned'.}

\SubjectFive reinforced this need, \myinlinequote{considering...that people are leaving quite frequently and their knowledge is lost...something like lessons learned would definitely be helpful.} 
When such documentation does exits, \SubjectFour recalls its usefulness, \myinlinequote{I have used what other team members have written [about issues]...it saved me a lot of time...especially in cases [where] more people are joining the team or...leaving.}

\subsubsection{Knowledge loss due to fragmented documentation} \label{sec:Theme 3.4}
In the absence of a centralized system, failure knowledge was scattered across multiple platforms, making them difficult to locate and as a result, inconsistently used and often overlooked.
\SubjectFive expressed this frustration, 
\myinlinequote{I didn't know where to find stuff...I didn't know what was documented in GitLab...in the wiki...on the team [channel], and what was discussed in which meeting...And people don't know that it's there and don't look at it. So it doesn't really help.}
\SubjectSeven similarly described information retrieval issues:
\myinlinequote{We lose [knowledge] right now...when you have a wiki with a lot of pages where you didn't find anything anymore, because the search is bad.}
\SubjectTwo similarly exclaims, \myinlinequote{If no one gives you the link, usually you cannot find it...[It's more] `Oh, I saw this before somewhere.'}
Others acknowledged that this lack of organization poses a risk of knowledge loss. 
As \SubjectTwo warned, \myinlinequote{This knowledge could be lost because they are not well documented and not well processed to transfer it...to say, `part of your project is to write lesson learned.'} 
\SubjectOne exclaims, \myinlinequote{How to document learnings...this is a hurdle. How do we document that?} and later identifies a need: \myinlinequote{We don't have a lessons learned page or database or something like that.}
Many subjects like \SubjectSix suggested, \myinlinequote{I think that there should be more a formal process on how to do that.}

\subsubsection{Transferability}

\added{Pilot study subjects shared challenges in learning from failures that mirrored those identified in the case study.
All pilot subjects cited lack of time and resources as the major barrier (cf.~\cref{sec:Theme 3.1}). 
\PilotSubjectOne explained, \myinlinequote{trying to learn from failures...is very difficult to do...when you're short staffed...and [we] don’t have time to sit down and write everything we’ve learned, because [we’re] so busy.}
The \PilotTeam similarly emphasized that \myinlinequote{time and priorities are constraints.}
\PilotSubjectOne expressed how this leads to knowledge loss during team turnover, \myinlinequote{it becomes a big problem where senior engineers' knowledge isn't captured anywhere because they're so short on time.}
\PilotSubjectOne notes that even when failure knowledge is documented, it is often inaccessible,\myinlinequote{you can search around...but it's hard to actually get stuff you can sit down and read and get [lessons] out of.}}

\section{Discussion and Future Work} \label{sec:Discussion}


\subsection{Learning from Failure: Are We Equipped?}

This study set out to understand how software engineers learn from failures, with the assumption that HROs, such as a national space research center, would exemplify best practices (cf.~\cref{sec:Case study setting}). 
We selected this case study precisely because we expected its teams to operationalize the structured, cyclic processes for capturing, disseminating, and applying failure knowledge that are prescribed in standards for high-reliability engineering~\cite{christiansonBecomingHighReliability2011}.
However, our findings challenge this assumption. 
\added{Prior work proposed structured techniques for gathering, documenting, and sharing lessons from failures, with limited guidance on applying them (cf.~\cref{sec:Background-SoftwareEngineering}). 
In contrast, despite operating in an HRO environment, our subjects described largely informal, ad-hoc, and inconsistent approaches to learning from software failures (cf.~\cref{sec:Theme 1}).}
Many participants acknowledged recurring failures that were not systematically addressed, with lessons learned often confined to individual memory, undocumented conversations, or informal mentorship (cf.~\cref{sec:Case study setting}).
This pattern extended across our pilot study as well, which included subjects from other HROs and sectors (cf. \cref{sec:Transfer - Theme 1},~\cref{sec:Transfer - Theme 2}).

\added{The challenges faced by this organization to learn from failures mirror barriers in conducting postmortems identified two decades ago~\cite{kasiPostMortemParadox2008}. The persistence of these long-standing obstacles, even in a mission-critical HRO, underscores how deeply entrenched such barriers are in software engineering practice.}
\added{These findings raise a foundational question: If even such mission-critical HROs lack structured learning-from-failure practices, where should we look for models of effective failure learning?}
\replaced{And a broader concern:}{These observations raise a broader concern:} even though failure analysis is widely recommended in both safety-critical systems and modern DevOps cultures, \emph{do software teams actually know when, why, and how to learn from failures effectively?}
Our results suggest that even in organizations theoretically aligned with the principles of continuous improvement and resilience, failure learning remains underdeveloped, reactive, and fragile.
This disconnect points to a pressing need: the software engineering field may require clearer structures, practices, and cultural reinforcements to support failure learning. 
Without these, even well-intentioned teams may repeat the same mistakes, not due to negligence, but because they lack processes or tools for turning failures into actionable knowledge.

\added{Future work could look to domains where incentives for avoiding recurring failures are more tightly coupled to financial, contractual, or regulatory obligations. 
For example, cloud infrastructure providers with stringent service level agreements~\cite{beyer2016site}, banking systems with financial repercussions and regulatory oversight~\cite{PrinciplesOperationalResilience2021}, or government agencies with centralized engineering governance~\cite{NASAProceduralRequirements, LessonsLearnedResources2024}, may offer more developed models for learning from failures.}

\subsection{What Could Broader Software Engineering Practices Teach this HRO?}

While our case site operates in a mission-critical domain, many of its failure reflection and documentation practices lag behind those seen in large IT organizations.
Postmortem practices at some of these organizations, such as structured templates~\cite{dingsoyrPostmortemReviewsPurpose2005, beyer2016site} and searchable incident databases~\cite{beyer2016site, NASAPublicLessons, johnsonSoftwareSupportIncident2000}, are well documented in practice and literature~\cite{beyer2016site, allspawBlamelessPostMortemsJust2012}.
In contrast, subjects in our study described \textit{ad hoc} gathering of lessons and fragmented documentation (cf.~\cref{sec:Theme 1}).

Our results suggest that HROs could benefit from adapting practices from IT organizations, particularly in how they document and track incidents. 
Structured postmortem templates, searchable failure databases, and lightweight regular reflection could offer scalable processes to support failure learning without introducing excessive bureaucracy~\cite{dingsoyrPostmortemReviewsPurpose2005, beyer2016site}\added{, while also laying the foundational structure needed for future AI-supported tools to access and leverage failure knowledge at scale (cf.~\cref{sec:Discussion-Future})}.
Formalizing mentorship systems, on-boarding materials, or guidelines for recurring failure modes, could further strengthen resilience against knowledge loss~\cite{beyer2016site}.

Notably, as both the case study and pilot interviews revealed, even when lessons are documented, their application and integration into the SDLC was inconsistent.
Similarly, even in large IT and DevOps organizations, where postmortems and tooling are sometimes adopted, it remains unclear whether these efforts lead to meaningful learning or sustained changes in their SDLC. 
\deleted{This raises an important question: \textit{How do, and how can, learning-from-failure translate into sustained improvements in the SDLC?}}
\added{This raises important questions: How do existing practices for learning from failures translate into improvements? And how can such practices be designed to ensure that lessons from failures lead to sustainable improvments?}
The feedback cycle from lessons to tangible \replaced{improvements}{process change} remains poorly understood and, consequently, poorly supported by researchers.

\subsection{Learning from HROs: Strengths \& Opportunities}

Despite gaps in structured failure learning, our results reveal informal  practices that are worth preserving and enhancing. 
Given constraints, such as lack of time and resources (cf.~\cref{sec:Theme 3}), subjects adapted to ad-hoc yet effective ways of gathering, sharing, and applying critical insights, particularly through informal discussions, senior-to-junior mentoring, and creating proven patterns (cf.~\cref{sec:Theme 1}).
Engineers, aware of the high-stakes nature of their work, developed habits of caution, defensive programming, and sought peer input, showing initiative even without formal process backing.
The broader software engineering community, especially startups or agile teams where heavy process may be impractical, could learn from these flexible, human-centered approaches. 
However, as our results show, such informal systems are vulnerable to turnover, memory loss, and siloing (cf.~\cref{sec:Theme 3}). 
Perhaps the challenge is not to replace them, but to understand: \textit{How can we augment them with lightweight structures that support consistency and institutional memory?}

\subsection{Future Directions} \label{sec:Discussion-Future}
Based on the current practices (~\cref{sec:Theme 1}), their flaws (~\cref{sec:Theme 2}), and their challenges (~\cref{sec:Theme 3}) in learning from failures \added{observed in our case study}, we propose \replaced{two}{three} opportunities for future research.

\added{First, more research is needed to evaluate and measure the value added by different failure learning practices, particularly the tradeoffs between lightweight ad-hoc methods and structured approaches. 
\textit{When is informal reflection enough?
When does structure pay off?
What contextual factors (\eg team size, mission type) make one approach preferable over another?}
We note that there are no established ways to measure the value or effectiveness of such practices, making this measurement gap itself an open research problem.
Additionally, all subjects repeatedly noted that time and resources are major constraints~\cite{kasiPostMortemParadox2008, desouzaExperiencesConductingProject2005c, vieiraTechnicalManagerialDifficulties2019,lyytinenLearningFailureInformation1999, dingsoyrPostmortemReviewsPurpose2005}, even when the cost of failures (\ie the value of failure learning) is acknowledged. 
Our data suggest the need to design and evaluate lightweight failure learning practices that support gathering, documentation, and reflection of lessons without imposing significant overhead. 
Although this need has been discussed in prior works~\cite{dingsoyrAugmentingExperienceReports2001,dingsoyrPostmortemReviewsPurpose2005,myllyaho2004review}, it appears to still be unresolved in practice.}

\replaced{Second}{Third}, automation tools may reduce the burden of capturing and retrieving failure knowledge. 
There are existing tools to help manage incidents~\cite{FireHydrantIncidentManagement, PagerDutyRealTimeOperations}, but they are limited help for leveraging lessons learned. 
Recent works have explored using LLMs to synthesize incident data into reports~\cite{anandayuvaraj2024fail, jinAssessSummarizeImprove2023}.
Combining such automated reports with Retrieval-Augmented Generation (RAG)~\cite{lewisRetrievalaugmentedGenerationKnowledgeintensive2020}, or other Generative Artificial Intelligence (GenAI)~\cite{alaviKnowledgeManagementPerspective2024} techniques, may also enable engineers to easily leverage organizational knowledge on failure data~\cite{ahmedRecommendingRootCauseMitigation2023a} and lessons. 
For example, this could enable a failure-aware chatbot to provide engineers with targeted, context-sensitive advice grounded in past failures. 
However, as our findings highlight, existing tools such as wikis and questionnaires are often underutilized, suggesting that automation efforts must integrate seamlessly into developers’ existing workflows to encourage adoption.

\section{Conclusion} \label{sec:Conclusion}
Learning from failures is widely regarded as essential for improving reliability in high-reliability organizations, however there is a knowledge gap on how it is done in practice. 
This study presents the first in-depth qualitative analysis of how software engineers in an HRO gather, document, share, and apply lessons from failures. 
Through interviews with 10 engineers at a space research center, supplemented by 5 pilot interviews at other HROs, we found that failure learning is informal, ad-hoc, and inconsistently integrated into software development practices, even in settings where reliability is critical.
Our findings reveal challenges including time and resource constraints, knowledge loss due to team turnover and fragmented documentation, and weak process enforcement. 
\replaced{Additionally,}{Although HROs and IT organizations are held as exemplars for structured postmortems,} the feedback loop from \replaced{lessons learned}{documented insights} to improvements remains poorly understood \deleted{across domains}.
Our findings shed light on the real-world practices and challenges for learning for software failures in HROs, and suggests future research directions to support a failure-aware software development life-cycle.

\section{Acknowledgments}
This work was supported by
  the German DAAD Rise fellowship,
  the German Aerospace Center (DLR),
  Cummins,
  the Qualcomm Innovation Fellowship, and 
  the Jet Propulsion Laboratory~\footnote{Portions of this work were performed by the Jet Propulsion Laboratory, California Institute of Technology, under a contract with the National Aeronautics and Space Administration (80NM0018D0004). Reference herein to any product, process, or service by trade name, trademark, or manufacturer, does not imply its endorsement by the United States Government or JPL.}

\ifARXIV
\else
\section{Data Availability} \label{sec:DataAvailability}
Our artifact (\url{https://doi.org/10.5281/zenodo.17754815})
contains our interview protocol, codes, and themes. 
We omit transcripts due to the difficulty in anonymizing a case study and the sensitive nature of the data.
\fi

\raggedbottom
\pagebreak

\bibliographystyle{bib/ACM-Reference-Format}
\bibliography{bib/PurdueDualityLab}

\ifAPPENDIX

\clearpage 
\appendix

\section*{Outline of Appendices}

The appendix contains the following material:
\begin{itemize}
\item \Cref{sec:appendix-InterviewProtocol}: The interview protocol. 
\item \Cref{sec:appendix-CodeCategorizedByThemes}: The full code book organized by themes \& subjects. 
\item \Cref{sec:appendix-CodebookEvolutionDiagram}: The evolution of the Codebook used in our analysis.
\item \Cref{sec:appendix-Demographics}: Additional demographics for the case study subjects.
\end{itemize}

\section{Interview Protocol} \label{sec:appendix-InterviewProtocol}
\cref{tab:InterviewProtocolSummary} gave a summary of the interview protocol.
Here we describe the full protocol in \cref{table:postmortem-interview}.

\noindent\cref{table:interview-rq} maps our interview protocol to our research questions.

\begin{table*}[ht]
\small
\centering
\caption{Interview Protocol for Postmortems in Software Failures}
\begin{tabularx}{\textwidth}{cX}
\toprule
\textbf{Category} & \textbf{Questions} \\
\toprule

\textbf{Reflective Example} & 
Q1: Can you provide an example of a time when your team studied a software/engineering issue to gather lessons? \\
& Q2: After the first time, did this issue reoccur? \\
& Q3: Were measures taken to prevent the issue from reoccurring? How? \\
& Q4: Did the issue reoccur after these measures? If so, what actions were taken afterward? \\
& Q5: If recurred: Can you provide an example of an issue with long-term remediation where the issue stopped recurring? \\
& Q6: If did not recur: Can you provide an example of an issue recurring despite initial fixes? \\
\midrule

\textbf{Postmortem Analysis} & Q7: How does your team study or handle software issues when they occur? \\
& Q8: What happens once the issue is fixed? \\
& Q9: In what scenarios are issues studied beyond just fixing the issue? \\
& Q10: Are these issues communicated amongst other team members or stakeholders? How? \\
& Q11: Would your team prioritize studying issues with impacts on humans or the environment? \\
& Q12: Do you have insights on how reflective practices differ for hardware or mechanical failures? \\
& Q13: Are there meetings to discuss lessons from issues? How are they structured? \\
& Q14: How are issues and lessons documented? How often and in what cases are they referenced? \\
& Q15: How are these issues and lessons stored (Internal wiki, Jira, GitLab, Confluence)? \\

\midrule
\textbf{Learning Integration} & Q16: How are lessons learned integrated into future product development? \\
& Q17: Are issues from past projects used in fault modeling during new projects? \\
& Q18: If not used, would it be beneficial to use past lessons? \\
& Q19: If so, how are they beneficial and effective? \\
& Q20: What changes resulted from studying past issues? \\
& Q21: How are issues and lessons shared across disciplines or teams? \\
& Q22: How are these lessons communicated to new team members (training, reference material)? \\

\midrule
\textbf{Processes} & Q23: Regarding activities you've described, how standardized are these processes? \\
& Q24: Are these standardized across all teams? Why or why not? \\
& Q25: Were these activities different in your past teams/organizations? How? \\
& Q26: Do you see gaps or opportunities for improvement in these processes? \\
& Q27: To what extent are activities automated, from documenting to learning from issues? \\
& Q28: What further tools or automation would you suggest to help document and learn from issues? \\

\midrule
\textbf{Benefits of Postmortems} & Q29: Textbooks and research literature say that studying and learning from failures is an important process. Do you believe that your team would/has benefit from such processes? Why? \\
& Q30: Do you think there would be changes in your products due to such processes? \\
& Q31: Would formal processes like reading clubs or role-playing exercises be useful? Why? \\
& Q32: Do you think these processes are practical? What constraints concern you? \\

\midrule
\textbf{Automation} & Q33: What type of automation would help document issues beyond just tracking them? \\
& Q34: What type of automation would help learn from issues? \\
& Q35: Would a database linking issues with lessons learned help your development process? \\
& Q36: How would a chatbot for querying project-specific issue information assist your team? \\

\bottomrule
\end{tabularx}
\label{table:postmortem-interview}
\end{table*}

\begin{table*}[ht]
\small
\centering
\caption{Interview Protocol organized by Research Questions (RQs). Some interview questions address more than one research question.}
\begin{tabularx}{\textwidth}{cX}
\toprule
\textbf{RQ} & \textbf{Questions} \\
\toprule

\textbf{RQ1: How do software teams gather lessons from failures?} & 
Q1--Q6 (Reflective examples of recurring or remediated issues) \\
& Q7--Q9 (How issues are studied and handled when they occur) \\
& Q13--Q15 (Meetings to discuss lessons; how lessons are documented and stored) \\

\midrule
\textbf{RQ2: How are these lessons shared \& integrated into the SDLC?} & 
Q10 (Communication with team members or stakeholders) \\
& Q16--Q22 (Integration into future projects, fault modeling, benefits, onboarding/training, cross-team sharing) \\

\midrule
\textbf{RQ3: What are the challenges in learning from failures?} & 
Q11--Q12 (Prioritization of issues; differences from hardware/mechanical failures) \\
& Q23--Q25 (Standardization across teams and organizations; comparison with past teams/orgs) \\
& Q26--Q32 (Gaps, opportunities for improvement, practicality, constraints, perceived benefits) \\
& Q33--Q36 (Automation and tooling challenges/opportunities) \\

\bottomrule
\end{tabularx}
\label{table:interview-rq}
\end{table*}

\section{Code Categorized by Themes} \label{sec:appendix-CodeCategorizedByThemes}

\Cref{tab:failure-codes-fullpage} presents the full set of codes identified through our qualitative analysis. We organized these codes into three overarching themes that emerged from the data: (1) failure learning is informal and inconsistent, (2) recurring failures and the illusion of effective failure management, and (3) challenges within the learning-from-failures process.

Within each theme, we group related codes into subthemes to highlight more specific patterns. For each code, we also indicate which subjects (S1--S10, PS1--PS4, PTeam) mentioned the issue during interviews. This structure allows us to both quantify the spread of each code and contextualize how failure-related practices differ across participants.

\renewcommand{\arraystretch}{1.5}
\begin{table*}[!t]
\centering
\caption{The codebook developed from and applied to our data, organized by themes and our subjects (outlined in~\cref{tab:Demographics}).}
\scriptsize
\begin{tabularx}{\textwidth}{>{\raggedright\arraybackslash}p{0.6\linewidth}X}
\toprule
\textbf{Code} & \textbf{Subjects} \\
\midrule

\textbf{Theme 1: Failure learning is informal \& inconsistent} & \\

\underline{Gathering lessons from failures is informal \& inconsistent} & \\
Example of \textit{ad hoc} discussion of failure & S2, S5, S6, S7, S10 \\
Failure discussion meeting structure is \textit{ad hoc} & S1, S2, S3, S4, S5, S6, S8, S9, S10, PS2, PS3, PTeam \\
There is a not a blaming culture, enabling open failure discussion & S3, S5, S6 \\
There is no process to learn from failures & S1, S2, S3, S4, S5, S6, S7, S8, S9, S10, PS1, PS3, PTeam \\
Studying, learning, sharing failure knowledge is \textit{ad hoc} & S1, S2, S4, S5, S6, S7, S8, S9, PS2, PTeam \\
Reflection of good, bad, and improvements \textit{ad hoc} & S3, S4, S7, S8, S10 \\
Lessons learnt are gather for project management but not software & S7, S9 \\
Lessons learnt is gathered \textit{ad hoc} & S2, S5, S6, S7, S8, S9, PS2 \\
Example of \textit{ad hoc} gathering of lessons learnt & S2, S3, S5, S6, S7 \\
Disconnect between subjects thinking that: discussion of failure = gathering lessons learnt & S2, S4, S6, S9, S10, PTeam \\
Disconnect: assumption that lessons learnt are gathered when they are not & S3, S4, S6, S7, S9, PS1, PS2, PS3, PTeam \\
Failure is discussed upto initial fix (~prevention or lessons learnt) & S1, S2, S3, S4, S5, S6, S7, S8, S9, S10, PS2, PS3 \\

\underline{Documentation of lessons learned is inconsistent \& fragmented} & \\
Example of \textit{ad hoc} documentation of failure & S2, S6 \\
Failure knowledge documentation practices are \textit{ad hoc} & S1, S2, S3, S4, S5, S6, S7, S10, PS1, PS2, PS3 \\
Methods to document failure knowledge & S2, S4, S5, S7, S8, PS1, PS2, PS3, PTeam \\
Failure is documented upto fix (~prevention or lessons learnt) & S1, S2, S3, S4, S6, S7, S8, S9, S10, PS1, PS3, PTeam \\
Lessons learnt is documented \textit{ad hoc} & S2, S6, S7, S8, S9, PTeam \\
Lessons learnt is NOT explicitly documented & S1, S2, S4, S5, S6, S8, S10, PS1 \\

\underline{Lessons learned are transferred \& applied informally} & \\
Failure knowledge is shared \textit{ad hoc} & S1, S2, S3, S4, S5, S6, S7, S8, S9, S10, PTeam \\
Past failure knowledge is provided to new members & S2, S4, PS1, PS2 \\
Past failure knowledge is NOT provided to new members & S1, S5, S9, PS3, PTeam \\
Senior members transfer internalized lessons \textit{ad hoc} & S1, S2, S3, S4, S5, S6, S7, S8, S9, S10, PS1 \\
Sharing is complicated with classified knowledge & S2, S5, S6, S7, S8, S10 \\
Lessons learnt is NOT standardly shared & S1, S2, S10, PTeam \\
Example of lessons learnt being applied in practice & S1, S3, S6, S7, S8, S9, S10, PS2, PTeam \\
Disconnect between past failure knowledge being referenced when it is not referenced often & S2, S3, S4, S7, S9, PS1 \\
Failure knowledge documentation is reviewed later & S2, S3, S4, S5, S8, S10, PS2, PTeam \\
Failure knowledge documentation is NOT reviewed later & S1, S2, S3, S6, S7, S8, S10, PS2, PS3 \\

\addlinespace
\midrule

\textbf{Theme 2: Recurring failures and the illusion of effective failure management} & \\
Disconnect between subjects thinking their failure management is working, except they report recurring issues & S2, S7, PS2 \\
Disconnect: assumption that general software practices (reusability, code quality, etc) can prevent recurring failures & S1, S3, S4, S6, S9, PS3 \\
There are recurring failures & S1, S4, S5, S6, S7, S9, PS3, PTeam \\
Example of failure not recurring & S2, S3, S6, S7 \\
Example of failure recurring & S1, S2, S3, S5, S6, S7, S8, S10, PS3 \\
Recurring failure due to inadequate failure documentation practices & S1, S2, S7 \\
Recurring failure due to not sharing & S2, S7 \\
Recurring failure due to fix for a specific case (Need generalizable processes, postmortems) & S2 \\
Disconnect: assumption that only new members need process to learn from failures & S7, PS2 \\

\addlinespace
\midrule

\textbf{Theme 3: Challenges within the learning-from-failures process} & \\
Formal process for lessons learnt is standardized, but is too general and not useful & S1, S7 \\
SDLC is different/informal due to research org & S1, S3, S4, S6, S8 \\
Loss of knowledge due to dynamic team changes & S2, S4, S5, S8, S9, PTeam \\
Loss of knowledge due to inadequate failure documentation practices & S2, S5, S7, S9, PS1 \\
Lack of time and resources to conduct traditional postmortems & S1, S2, S3, S4, S5, S6, S7, S8, S9, S10, PS1, PS3, PTeam \\
Inclination to have a process to learn from failures if impact was on physical world & S3, S5, PS1, PS3 \\
More effort to prevent failures in traditional engineering at earlier dev phase & S1, S4, S5, S6, S7, S9, PTeam \\

\bottomrule
\end{tabularx}
\label{tab:failure-codes-fullpage}
\end{table*}

\section{Code Evolution Diagram} \label{sec:appendix-CodebookEvolutionDiagram}

\cref{fig: evolution-fig} describes the evolution of our codebook over six rounds of revisions.

\begin{figure*}[t]
    \centering
    \includegraphics[width=0.77\linewidth]{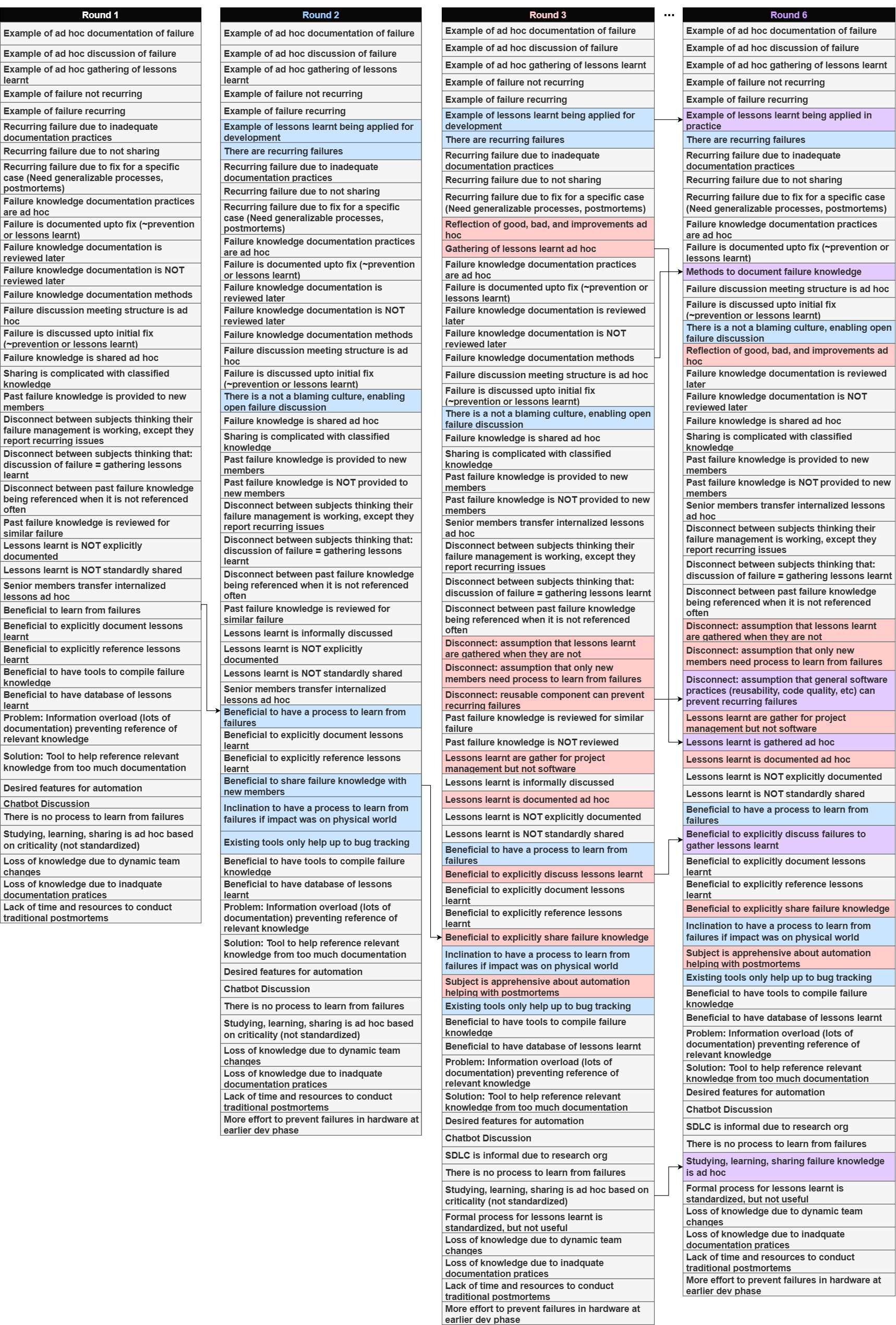}
    \caption{ Codebook Evolution: The codebook underwent six rounds of revisions, with changes highlighted by different colors: round 2 (blue), round 3 (red), and round 6 (purple). Arrows indicate the replacement of codes. The final codes from round 6 were used in this paper, and categorized thematically.
    }
    \label{fig: evolution-fig}
\end{figure*}

\section{Additional Demographics for Case Study Subjects} \label{sec:appendix-Demographics}

\cref{fig:EducationLevel} illustrates the highest degree held by participants.

\begin{figure}[ht]
    \centering
    \includegraphics[width=1\linewidth]{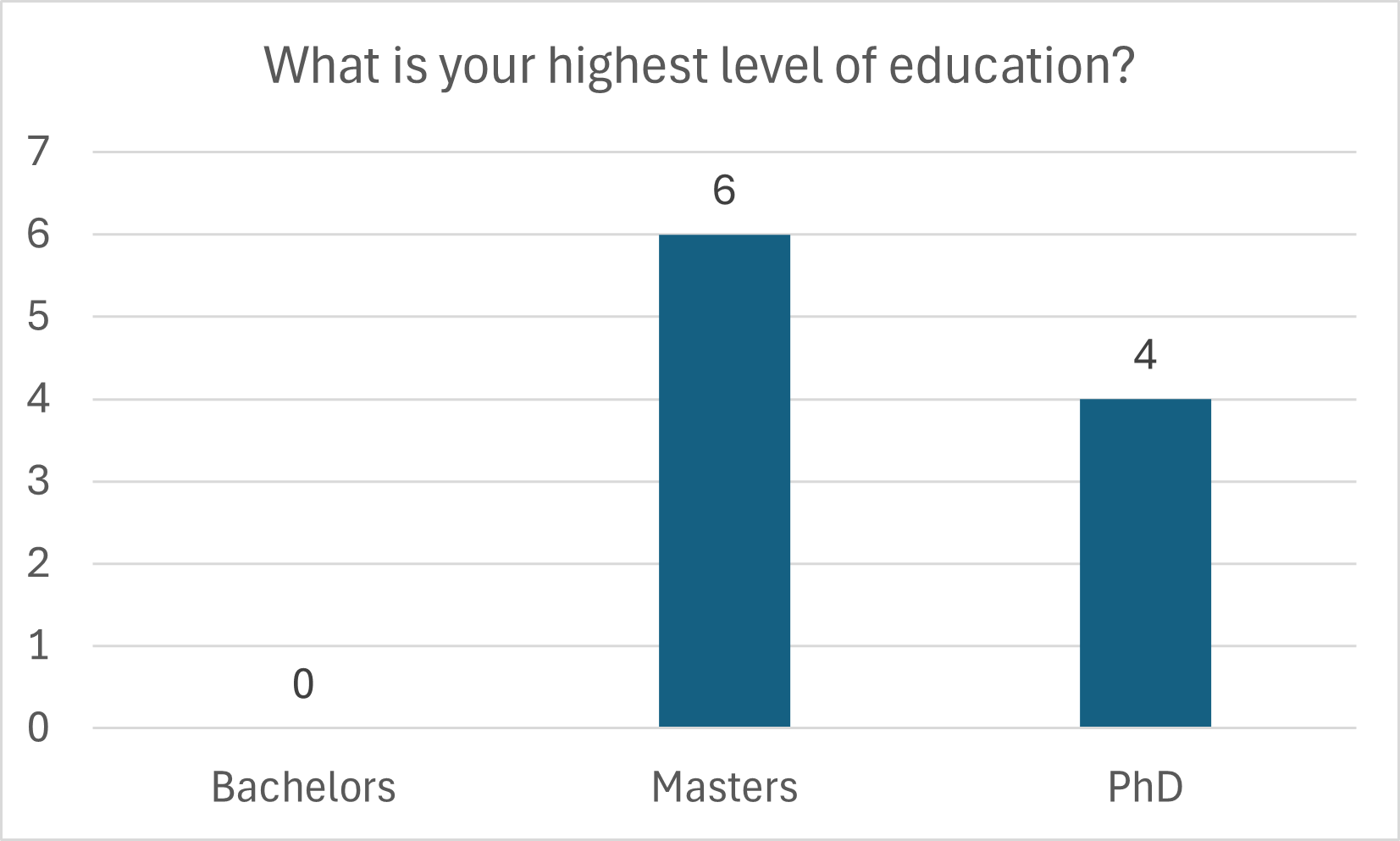}
    \caption{Highest degree held by participants.}
    \label{fig:EducationLevel}
\end{figure}

\noindent\cref{fig:Degree} illustrates the disciplinary background of participants.

\begin{figure}[ht]
    \centering
    \includegraphics[width=1\linewidth]{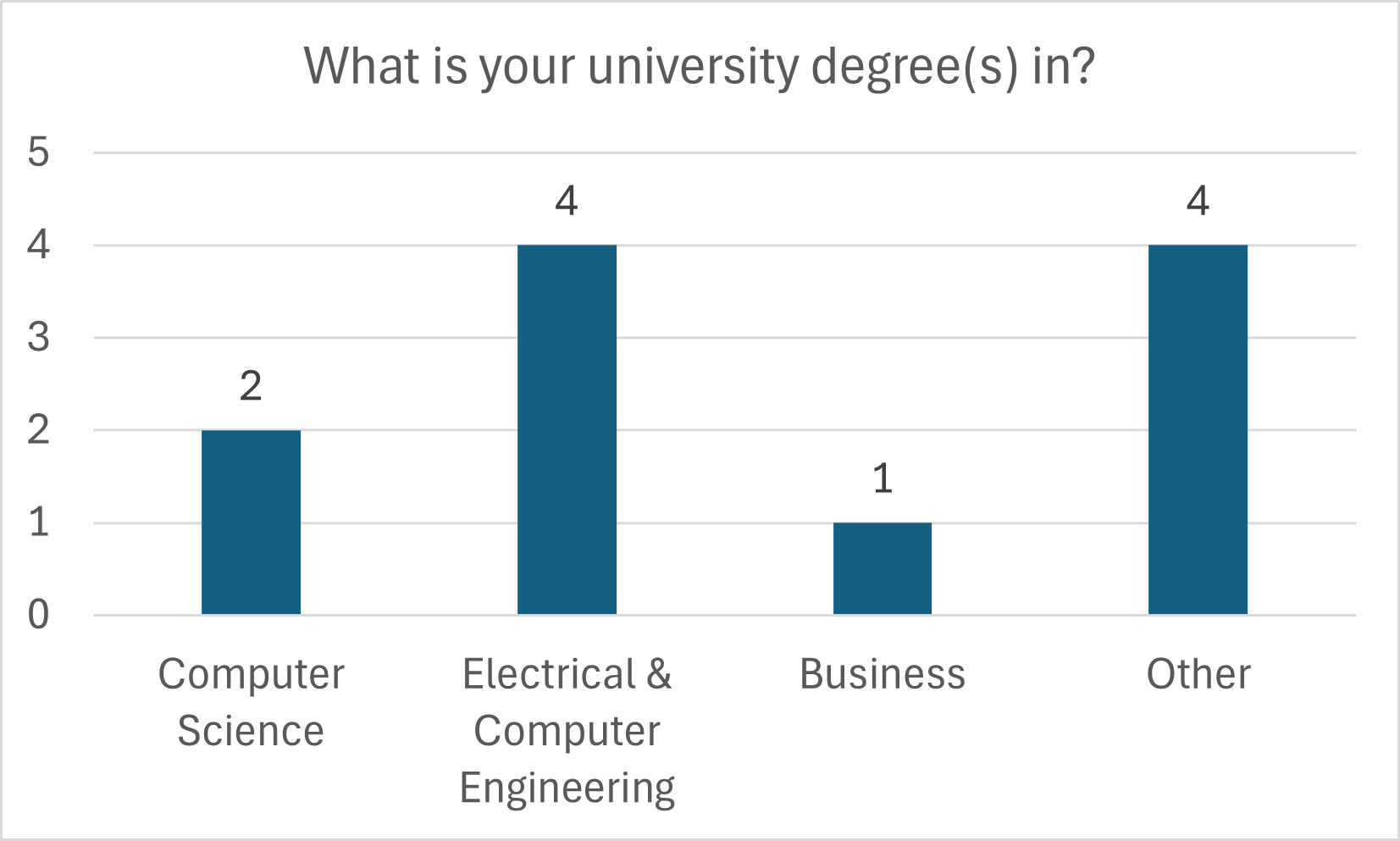}
    \caption{Disciplinary background of participants.}
    \label{fig:Degree}
\end{figure}

\noindent\cref{fig:Training} illustrates where participants reported learning about software engineering practices.

\begin{figure}[ht]
    \centering
    \includegraphics[width=1\linewidth]{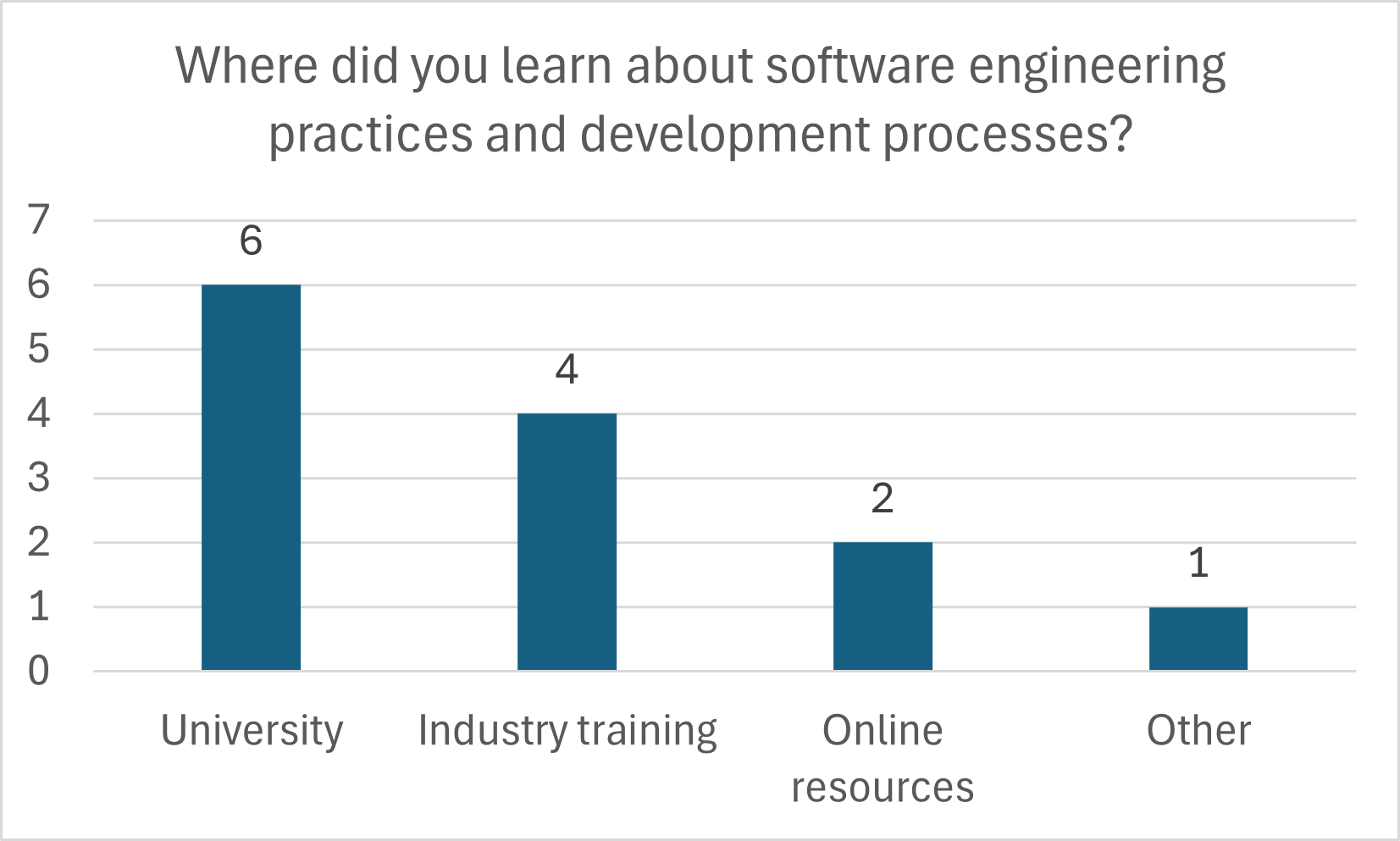}
    \caption{Where participants reported learning about software engineering practices.}
    \label{fig:Training}
\end{figure}

\clearpage

\fi 

\end{document}